\documentclass[preprint,showpacs,showkeys,amsmath,amssymb,11pt,a4paper]{article}
\pdfoutput=1
\usepackage[utf8]{inputenc}
\usepackage{jheppub}
\title{Phenomenology of two texture zero neutrino mass in left-right symmetric model with $Z_8 \times Z_2$}
\author[a]{Happy Borgohain,}
\author[b]{Mrinal Kumar Das%
}

\affiliation[a,b]{Department of Physics, Tezpur University, Tezpur 784028, India}

\emailAdd{happy@tezu.ernet.in}
\emailAdd{mkdas@tezu.ernet.in}

\abstract{
	We have done a phenomenological study on the neutrino mass matrix $M_\nu$ favoring two zero texture in the framework of left-right symmetric model  (LRSM) where type I and type II seesaw naturally occurs. The type I SS mass term is considered to be following a trimaximal mixing (TM) pattern. The symmetry realizations of these texture zero structures has been realized using the discrete cyclic abelian  $Z8\times Z2$ group in LRSM. We have studied six of the popular texture zero classes named as A1, A2, B1, B2, B3 and B4 favoured by neutrino oscillation data in our analysis. We basically focused on the implications of these texture zero mass matrices in low energy phenomenon like neutrinoless double beta decay (NDBD) and lepton flavour violation (LFV) in LRSM scenario. For NDBD, we have considered only the dominant new physics contribution coming from the diagrams containing purely RH current and another from the charged Higgs scalar while ignoring the contributions coming from the left-right gauge boson mixing and heavy light neutrino mixing. The mass of the extra gauge bosons and scalars has been considered to be of the order of TeV scale which is accessible at the colliders.  }
\keywords{Discrete flavour symmetry, neutrino physics, spontaneous symmetry breaking}

\begin{document}

 \maketitle

	\section{INTRODUCTION}

	With the landmark discovery of neutrino oscillation and corresponding realization that neutrinos are massive and they mix during propagation have brought into limelight several interesting consequences like necessity of going beyond the succesful standard model (SM).
	Global analysis of neutrino oscillation data has quite precisely determined the best fit and 3$\sigma$ ranges of neutrino parameters, viz., the mixing angles, mass squared differences, the Dirac CP phase $\delta$ \cite{de2018status}, but the absolute neutrino mass and the additional CP phase (for Majorana particles) $\alpha$ and $\beta$ are not accurately found yet. Nevertheless several other questions are yet not perceived amongst which notable is understanding the origin and dynamics of the neutrino mass and the lepton flavour structures of the fermions. The role of symmetry in particle physics \cite{ishimori2010non} cannot be overestimated. It is utmost important to understand the underlying symmetry inorder to understand the origin of neutrino mass and the leptonic mixing. Symmetries can relate two or more free parameters or can make them vanish, thereby making the model more predictive. One of the possible role flavour symmetry can play is to impose texture zeros \citep{ludl2014complete, frampton2002zeroes,xing2002texture, singh2016revisiting, ahuja2004texture} in the mass matrix and to reduce the number of free parameters.  For a symmetric $M_\nu$, it has six independent complex entries. If n of them are considered to be vanishing, we arrive at $^6C_n=\frac{6!}{n!(6-n)!}$ different textures. A texture of $n \geqslant 3$ is not compatible with current experimental data and neutrino mixing angles. Texture zero approaches has been established as a feasible framework for explaining the fermion masses and mixing data in quark as well as lepton sector and has been studied in details in a large number of past works like \cite{meloni2014two, borah2015discriminating,fritzsch2011two, alcaide2018fitting,zhou2016update, kaneko2003neutrino, dev2016two, nath2017phenomenological,agarwalla2018addressing}. Specifically two texture zero mass matrices are considered to be more interesting as they can reduce maximum number of free parameters. Two independent zeroes in the matrix can lead to four relations among the nine free parameters in the neutrino mass matrix which can be checked against the available
	experimental data.

	In the simplest case one can presume the charged lepton mass matrix to be diagonal and then consider the possible texture zeros in the symmetric Majorana mass matrix. Considering the basis in which charged lepton mass matrix is diagonal there are different categories of two zero texture neutrino mass matrix out of which some are ruled out and some are marginally allowed. Glashow et al.\cite{frampton2002zeroes} have found seven acceptable textures of neutrino mass matrix (out of total fifteen)  with two independent vanishing entries in the flavour basis for a diagonal charged lepton mass matrix to be consistent with current experimental data.

	\par
	
Neutrino mass and mixing matrix have different forms based upon some flavour symmetries. Amongst them, the most popular one which is consistent with neutrino oscillation data is the Tribimaximal mixing (TBM) \cite{harrison2002tri} structure as proposed by Harison, Perkins and Scott. The resulting mass matrix in the basis of a diagonal charged lepton mass matrix is both 2-3 symmetric and magic. By magic, it means the row sums and column sums are all identical. The reactor mixing angle $\theta_{13}$ vanishes in TBM because of the bimaximal character of the third mass Eigen state $\nu_3$.
	 However  $\theta_{13}$ has been measured to be non zero by experiments like T2K, Daya Bay, RENO and DOUBLE CHOOZ \cite{abe2014observation,an2012observation,ahn2012observation,ahn22012observation}, which demands for a correction to the TBM form which may be a correction or some perturbation to this type. Henceforth, owing to the current scenerio of neutrino oscillation parameters several new models has been theorized and studied by the scientific communities. Amongst several neutrino mass models, Trimaximal mixing (TM) \cite{rodejohann2017trimaximal,luhn2013trimaximal,antusch2012trimaximal,kumar2010unitarity,albright2009comparing,grimus2008model,gautam2018trimaximal,gautam2016zeros} is one in which non zero reactor mixing angle can be realized. The mixing matrix consists of identical second column elements similar to the TBM type. However, it relaxes some of the TBM assumptions, since it allows for a non zero $\theta_{13}$ as well as preserves the solar mixing angle prediction. It will be discussed in details in the section III. Besides the zeros in the neutrino mass matrix which is one of currently studied approaches for precisely explaining neutrino masses and mixing can also be examined using the TM mixing.
	
	\par Inspite of enormous success, there are several unperceived problems in the neutrino sector which includes the absolute scale of neutrino mass, the mass hierarchy, the CP violation, the intrinsic nature of neutrinos, whether Dirac or Majorana. One of the important process which undoubtedly establish the Majorana nature of neutrinos (violation of lepton number by two units) is neutrinoless double beta decay (NDBD) (for a review see\cite{dell2016neutrinoless}). Besides, the observation of NDBD would also throw light on the absolute scale of neutrino mass and in explaining the matter, anti-matter asymmetry of the universe. The NDBD experiments like KamLAND-Zen, GERDA, EXO-200 \cite{shirai2017results,agostini2018upgrade,albert2018search} directly measures and provides bounds on the decay half life which can be converted to the effective neutrino mass parameter, $m_{ee}$ with certain uncertainity which arises due to the theoretical uncertainity in the NME. The current best limits on the effective mass $<m_{ee}>$ are of the order of 100 meV. The next generation experiments targets to increase the sensitivity in the 10 meV mass range. Thus, the future NDBD experiments can shed lights on several issues in the neutrino sector. Observing this rare decay process with the current experiments would signify new physics contributions beyond the standard model (SM) other than the standard light neutrino contribution.
	
		\par There are several BSM frameworks , amongst which one of the most fascinating and modest frameworks in which neutrino mass and other unsolved queries  can be addressed is the left right symmetric model (LRSM) \cite{senjanovic1975exact,mohapatra1980local,marshak1981majorana} where the gauge group is
	$SU(2)_L\times SU(2)_R\times U(1)_{B-L}$. It has become a topic of interest since long back owing to its indomitable importance and has been studied in details by several groups in different contexts \cite{abbas2008neutrino,dev2018displaced,ge2015new,barry2013lepton,awasthi2016implications,chakrabortty2012neutrinoless,borah2015neutrinoless,patra2013neutrinoless,parida2013left,bambhaniya2016scalar,nemevvsek2013connecting}. Herein the type I and type II seesaw arises naturally rather than by hand. The neutrino mass in LRSM can be written as a combination of both the type I and type II seesaw mass terms.
	A brief review of the LRSM has been presented in the next section. As far as NDBD is concerned, LRSM can give rise to several new physics (non standard) contributions coming from LH, RH, mixed, scalar triplet etc. Several analysis has been done already involving NDBD in LRSM \cite{barry2013lepton,borah2015neutrinoless,awasthi2016implications,chakrabortty2012neutrinoless,borah2015neutrinoless,patra2013neutrinoless,parida2013left,bambhaniya2016scalar} and their compatibility with LHC experiments \cite{ruiz2017lepton,maiezza2015lepton,deppisch2015probing,helo2014heavy,awasthi2016implications}.

	As cited by several authors \cite{gautam2016zeros,gautam2018trimaximal}, the TM mixing can satisfy the current neutrino experimental data when combined with two zero textures. In this context, we have done a phenomenological study on the neutrino mass matrix $M_\nu$ favouring two zero texture in the framework of LRSM where type I and type II seesaw naturally occurs. The type I SS mass term is considered to be following a TM mixing pattern. The symmetry realizations of these texture zero structures has been realized using the discrete cyclic abelian ($Z8 \times Z2$) group in LRSM. In order to obtain the desired two zero textures of the mass matrices, we have added two more LH and RH scalar triplets each. In our analysis, we have studied for the popular 6 texture zero classes being named as A1-A2  and B1-B4. We basically  focused in the implications of these texture zero mass matrices in low energy phenomenon like NDBD and LFV in LRSM scenerio. For NDBD, we have considered only the dominant new physics contribution  coming from the diagrams containing purely RH
	current mediated by the heavy gauge boson, $ \rm W_R$ by the exchange of heavy right handed neutrino, $ \rm N_R$ and another from the charged Higgs scalars
	mediated by the heavy gauge boson $ \rm W_R$  ignoring the contributions coming from the left-right gauge boson mixing and heavy light neutrino mixing as in our previous work \cite{borgohain2017neutrinoless}. The mass of the extra gauge bosons and scalars has been considered to be of the order of TeV accessible at the colliders.

	\par The paper has been organized as follows, in the next section we briefly review the LRSM, its particle contents along with texture zero and TM mixing. In section III we present the symmetry realizations of these classes by using a cyclic Z8$\times$Z2 group symmetry with possible particle contents
	in LRSM to obtain the desired texture zero matrices. Then in section IV we discuss NDBD and LFV in the framework of LRSM which is followed by the numerical analysis and results with the collider signatures in section V. We give the conclusion in section VI.

\section{MINIMAL LEFT-RIGHT SYMMETRIC MODEL AND TWO ZERO TEXTURE NEUTRINO MASS}

As has been mentioned before, LRSM is based on the gauge group $ \rm SU(3)_c\times SU(2)_L\times SU(2)_R\times U(1)_{B-L}$ \cite{senjanovic1975exact,mohapatra1980local,marshak1981majorana}, a simple extension of the standard model gauge group where parity is conserved at a very high scale . The spontaneous breaking of the left right symmetry then ensures violation of parity as observed at low energy scales. The usual type I and II seesaw are a necessary part of LRSM. The RH neutrinos are a necessary part of LRSM which acquires a Majorana mass when the
$SU(2)_R$ symmetry is broken at a scale $v_R$.\\
\subsection{PARTICLE CONTENTS:}
\begin{equation}\label{eqx}
Q^{'}_{L,R}=\left[\begin{array}{cc}
u^{'}\\
d^{'}
\end{array}\right]_{L,R}, \Psi^{'}_{L,R}=\left[\begin{array}{cc}
\nu_l\\
l
\end{array}\right]_{L,R},
\end{equation}
which are the quarks and leptons under LRSM
where the quarks are assigned with quantum numbers $(3,2,1,1/3)$ and $(3,1,2,1/3)$ and leptons with $(1,2,1,-1)$ and $(1,1,2,-1)$ respectively under
$ \rm SU(3)_c\times SU(2)_L\times SU(2)_R\times U(1)_{B-L}$.
 The Higgs sector in LRSM consists of the following multiplets,
\begin{equation}\label{eqx3}
\phi=\left[\begin{array}{cc}
\phi_1^0 & \phi_1^+\\
\phi_2^- & \phi_2^0
\end{array}\right]\equiv \left( \phi_1,\widetilde{\phi_2}\right), \Delta_{L,R}=\left[\begin{array}{cc}
{\delta_\frac{L,R}{\sqrt{2}}}^+ & \delta_{L,R}^{++}\\
\delta_{L,R}^0 & -{\delta_\frac{L,R}{\sqrt{2}}}^+ .
\end{array}\right].
\end{equation}
A bi-doublet with quantum number $ \rm \phi(1,2,2,0)$
and the $SU(2)_{L,R}$ triplets, $\Delta_L(1,3,1,2)$, $\Delta_R(1,1,3,2)$.The successive spontaneous symmetry breaking occurs as,
$\rm SU(2)_L\times SU(2)_R\times U(1)_{B-L}\xrightarrow{<\Delta_R>} SU(2)_L\times U(1)_Y \xrightarrow{<\phi>} U(1)_{em}$. The neutral component of the Higgs fields obtains a vacuum expectation value (vev), $<\delta_{R}^0>= v_R $, $<\delta_{L}^0>= v_L$, $<\phi_1^0>= k_1$ and
$<\phi_2^0>= k_2$ thereby providing masses for the the extra gauge bosons $ \rm (W_R$ and Z$ \rm \ensuremath{'})$ and for right handed neutrino field $ \rm (\nu_R)$,$ \rm W_L$ and Z bosons, Dirac masses for the quarks and leptons respectively.
 The vev of $\Delta_L$, $v_L$ plays a significant role in the SS relation which is the characteristics of the LRSM and can be written as, $<\Delta_L>=v_L=\frac{\gamma k^2}{v_R}$.

\par The Yukawa Lagrangian in the lepton sector is given by,
\begin{equation}\label{eqx8}
\mathcal{L}=h_{ij}\overline{\Psi}_{L,i}\phi\Psi_{R,j}+\widetilde{h_{ij}}\overline{\Psi}_{L,i}\widetilde{\phi}\Psi_{R,j}+f_{L,ij}{\Psi_{L,i}}^TCi\sigma_2\Delta_L\Psi_{L,j}+f_{R,ij}{\Psi_{R,i}}^TCi\sigma_2\Delta_R\Psi_{R,j}+h.c,
\end{equation}

 where,the indices $i,j=1,2,3$ represents the three generations of fermions. $C=i\gamma_2\gamma_0$ is the charge conjugation
operator, $\widetilde{\phi}=\tau_2\phi^*\tau_2$ and $\gamma_{\mu}$ are the Dirac matrices. Considering discrete parity symmetry, the Majorana Yukawa couplings $f_L=f_R$.
Equation (\ref{eqx8}) leads to $6\times6$ neutrino mass matrix as,

\begin{equation}\label{eqx8}
M_\nu=\left[\begin{array}{cc}
M_{LL}&M_D\\
{M_D}^T&M_{RR}
\end{array}\right],
\end{equation}

where
\begin{equation}\label{eqx9}
M_D=\frac{1}{\sqrt{2}}(k_1h+k_2\widetilde{h}), M_{LL}=\sqrt{2}v_Lf_L, M_{RR}=\sqrt{2}v_Rf_R,
\end{equation}

$M_D$, $M_{LL}$ and $M_{RR}$ being the Dirac neutrino mass matrix, left handed and right handed
Majorana mass matrix respectively. Assuming $M_L\ll M_D\ll M_R$, the
light neutrino mass, generated within a type I+II seesaw can be written as,

\begin{equation}\label{eq25}
\rm M_\nu= {M_\nu}^{I}+{M_\nu}^{II},
\end{equation}

\begin{equation}\label{eq26}
\rm M_\nu=M_{LL}+M_D{M_{RR}}^{-1}{M_D}^T
=\sqrt{2}v_Lf_L+\frac{k^2}{\sqrt{2}v_R}h_D{f_R}^{-1}{h_D}^T.
\end{equation}

Where the first and second terms in equation (\ref{eq26}) corresponds to type II seesaw and type I seesaw respectively.
Here,
\begin{equation}\label{eq27}
\rm h_D=\frac{(k_1h+k_2\widetilde{h})}{\sqrt{2}k} , k=\sqrt{\left|{k_1}\right|^2+\left|{k_2}\right|^2}
\end{equation}

In \cite{deshpande1991left}, authors introduced a dimensionless parameter $\gamma$ as,

\begin{equation}\label{eqx15}
\gamma=\frac{\beta_1k_1k_2+\beta_2{k_1}^2+\beta_3{k_2}^2}{(2\rho_1-\rho_3)k^2}.
\end{equation}

with the terms $\beta$, $\rho$ being the Higgs potential parameters.
The VEV for left handed triplet $v_L$ can be then be written as,

\begin{equation}\label{eq28}
\rm v_L=\frac{\gamma {M_W}^2}{v_R},  M_W=\frac{g k}{2}
\end{equation}

Both type I and type II seesaw terms can be written in terms of $M_{RR}$ in LRSM. Thus equation (\ref{eq26}) can be written as ,

\begin{equation}\label{eqx14}
M_\nu=\gamma(\frac{M_W}{v_R})^2M_{RR}+M_D{M_{RR}}^{-1}{M_D}^T.
\end{equation}

\subsection{TWO ZERO TEXTURE AND TM MIXING}

Non vanishing $\theta_{13}$ excluded $\mu-\tau$ symmetry to be an exact symmetry of the neutrino mass matrix which opts for a perturbation in $\mu-\tau$ symmetric mass matrix or a different form which gives rise to non zero $\theta_{13}$. Going through literature, we have seen that another form of symmetry known as magic symmetry can serve the purpose \cite{lam2006magic}. The corresponding mass matrix known as magic symmetric mass matrix can be made more predictive by imposing certain constraints in it. Adding zeroes in certain elements of the matrix can make it more anticipating. Certain types of one zero and two zero textures in neutrino mass matrix are consistent with neutrino data. We here study two zero texture in neutrino mass matrix $M_\nu$ which was first considered in\cite{frampton2002zeroes,xing2002texture} and subsequently by several other groups \cite{singh2016exploring,alcaide2018fitting,fritzsch2011two,zhou2016update,dev2007cp,singh2016revisiting,ahuja2004texture,meloni2014two,kaneko2003neutrino,dev2016two}. Trimaximal mixing (TM) in two texture zero has been extensively studied in literature \cite{gautam2016zeros,gautam2018trimaximal}. In TM, $\mu-\tau$ symmetry is broken but the magic symmetry is kept intact. It has been again named as $ \rm TM_1$ or $\rm TM_2$ based upon whether the second or the first column of the TBM mixing matrix remains intact respectively.  We have studied these allowed texture zeros in the magic neutrino mass matrix (satisfying $\rm TM_2$ mixing) which is the type I SS mass term in our case and studied its implications for low energy processes like NDBD and LFV. Two zero textures ensures two independent vanishing entries in the neutrino mass matrix. There are a total of $^6C_2$ i.e., 15 texture zeros of M$\nu$ which has been further classified into 6 sub
categories and can be named as- A1, A2; B1, B2, B3, B4; C1; D1, D2; E1, E2, E3; F1, F2, F3.
Out of the above, E1-E3; F1-F3 were ruled out, D1,D2 are marginally allowed and now
has been experimentally ruled out at 3$\sigma$ level. We are only left with 7 allowed cases of 2 zero
textures, viz., A1-A2; B1-B4 and C1, we are concerned with six of the above classes which are of the form ,

\begin{equation}
A1=\left[\begin{array}{ccc}
0 & 0 & X\\
0& X& X\\
X & X& X
\end{array}\right],A2=\left[\begin{array}{ccc}
0 & X &0 \\
X & X& X\\
0& X& X
\end{array}\right]
\end{equation}

\begin{equation}
B1=\left[\begin{array}{ccc}
X & X &0\\
X& 0& X\\
0 & X& X
\end{array}\right],B2=\left[\begin{array}{ccc}
X & 0&X \\
0 & X& X\\
0& X& 0
\end{array}\right], B3=\left[\begin{array}{ccc}
X & 0& X \\
0 & 0 & X\\
X & X & X
\end{array}\right], B4=\left[\begin{array}{ccc}
X & X& 0 \\
X & X & X\\
0 & X & 0
\end{array}\right]
\end{equation}

The neutrino mass matrix is said to be invariant under a magic symmetry and the corresponding mixing symmetry is known as trimaximal mixing (TM) with the $\rm TM_2$ mixing matrix given by (cite),

\begin{equation}\label{y1}
{U_{TM}}_2=\left[\begin{array}{ccc}
\sqrt{\frac{2}{3}} cos \theta& \frac{1}{\sqrt{3}}  & \sqrt{\frac{2}{3}} sin \theta\\
-\frac{cos \theta}{\sqrt{6}}+\frac{e^{-i\phi}sin\theta}{\sqrt{2}}& \frac{1}{\sqrt{3}}& 	-\frac{sin \theta}{\sqrt{6}}-\frac{e^{-i\phi}cos\theta}{\sqrt{2}}\\
-\frac{cos \theta}{\sqrt{6}}-\frac{e^{-i\phi}sin\theta}{\sqrt{2}} & \frac{1}{\sqrt{3}}& 	-\frac{sin \theta}{\sqrt{6}}+\frac{e^{-i\phi}cos\theta}{\sqrt{2}}
\end{array}\right],
\end{equation}
where $\theta$ and $\phi$ being the free parameters. It diagonalizes the magic neutrino mass matrix, which can be parameterized as,

\begin{equation}
M_{magic}=\left[\begin{array}{ccc}
p & q & r\\
q& r& p+r-s\\
r & p+r-s& q-r+s
\end{array}\right]
\end{equation}
 The different allowed classes of two zero texture along with their respective constraint equations are as shown below:
\begin{table}[h!]
	\centering
	\begin{tabular}{||c| c| c||}
		\hline
		$Class$ & Constraint equations \\ \hline
		$A_1$& $M_{ee}=0, M_{e\mu}=0$ \\ \hline
		$A_2$ &$M_{ee}=0, M_{e\tau}=0$  \\ \hline
		$B_1$ & $M_{e\tau}=0, M_{\mu\mu}=0 $\\ \hline
		$B_2$&$M_{e\mu}=0, M_{\tau\tau}=0$    \\ \hline
		$B_3$&$M_{e\mu}=0, M_{\mu\mu}=0$  \\ \hline
		$B_4$&$M_{\mu\mu}=0, M_{\tau\tau}=0$  \\ \hline
	\end{tabular}
\end{table}
Using these constraint equations, we can arrive at the different classes of two zero textured neutrino mass matrix favouring $\rm TM_2$ mixing.

\section{SYMMETRY REALIZATIONS IN LRSM}{\label{sec:level5}}

Several earlier works \cite{singh2016exploring,alcaide2018fitting,fritzsch2011two,zhou2016update,dev2007cp,singh2016revisiting,ahuja2004texture,meloni2014two,kaneko2003neutrino,dev2016two} has explained two zero texture which has been explored beyond standard model to address neutrino masses and mixing. In this work, we have extended the minimal left-right symmetric model by introducing two more left handed and right handed scalar triplets represented by ${\Delta_L}^{'}, {\Delta_L}^{''}$  and ${\Delta_R}^{'}, {\Delta_R}^{''}$ respectively to realize the desired textures of Dirac and Majorana mass matrix, $M_D$ and $M_{RR}$ while keeping in mind that the charged lepton mass matrix is diagonal. The symmetry realizations of these texture zero structures has been worked out using the discrete abelian ($Z_8 \times Z_2$) group in the framework of LRSM which are explained below.

\textbf{Class A1:}

The symmetry realization for the class A1 is shown in tabular form as below,

\begin{table}[h!]
	\centering
	\begin{tabular}{||c| c| c| c| c| c| c| c||}
		\hline
		$l_L$ & $Z_8 \times Z_2$ & $l_R$&$Z_8 \times Z_2$ &Higgs(LH)&$Z_8 \times Z_2$&Higgs(RH)&$Z_8 \times Z_2$\\ \hline
		$l_{Le}$& $(\omega^{6},-1)$ &  $l_{Re}$&$(\omega^{2},-1)$&$\Delta_L$&$(\omega^{7},1)$&$\Delta_R$&$(\omega,1)$\\ \hline
		$l_{L\mu}$ &$(\omega^{3},1)$  &  $l_{R\mu}$&$(\omega^{5},1)$&${\Delta_L}^{'}$&$(\omega^{2},1)$&${\Delta_R}^{'}$&$(\omega^{6},1)$\\ \hline
		$l_{L\tau}$ & $(\omega^{3},-1)$&  $l_{R\tau}$&$(\omega^{5},-1)$&${\Delta_L}^{''} $&$ (\omega^{2},-1),   $&${\Delta_R}^{''}$&$(\omega^{6},-1)$\\ \hline

	\end{tabular}
	\caption{Particle assignments for A1} \label{t1}
\end{table}

In all the classes the bidoublets $\phi$ and $\tilde{\phi}$ transforms as singlets $(1\times 1)$
 under the cyclic group $Z_8\times Z_2$. The diagonal Dirac and the charged lepton mass term(which is same for all the cases), in the matrix form can be written as,
\begin{equation}
M_D=\left[\begin{array}{ccc}
1 & \omega^3& \omega^3 \\ \omega^5 & 1 & -1\\
\omega^5 & -1 & 1
\end{array}\right]+\left[\begin{array}{ccc}
1 & \omega^3& \omega^3 \\ \omega^5 & 1 & -1\\
\omega^5 & -1 & 1
\end{array}\right]=2\left[\begin{array}{ccc}
1 & \omega^3& \omega^3 \\ \omega^5 & 1 & -1\\
\omega^5 & -1 & 1
\end{array}\right]\simeq \left[\begin{array}{ccc}
\times & 0& 0 \\ 0 & \times & 0\\
0 & 0& \times
\end{array}\right]
\end{equation}

 The corresponding Dirac Yukawa Lagrangian for all the cases can be written as,

\begin{equation}\label{eqx8}
\mathcal{L_D}=Y_{ee}\overline{L}_{Le}\phi\overline{L}_{Re}+\tilde{Y_{ee}}\overline{L}_{Le}\tilde{\phi}\overline{L}_{Re}+
Y_{\mu\mu}\overline{L}_{L\mu}\phi\overline{L}_{R\mu}+\tilde{Y_{\mu\mu}}\overline{L}_{L\mu}\tilde{\phi}\overline{L}_{R\mu}+
Y_{\tau\tau}\overline{L}_{L\tau}\phi\overline{L}_{R\tau}+\tilde{Y_{\tau\tau}}\overline{L}_{L\tau}\tilde{\phi}\overline{L}_{R\tau}
\end{equation}

Under these symmetry realizations, we get the Majorana mass terms (LH and RH) and the type I SS mass terms for the class A1, in the matrix form  as,
\begin{equation}
M_{RR}=\left[\begin{array}{ccc}
\omega^5 & -1& 1\\ -1 & 1 & 1\\
1 & 1 & 1
\end{array}\right],M_{LL}=\left[\begin{array}{ccc}
\omega^3 & -1& 1 \\ -1 & 1 & 1\\
1 & 1 & 1
\end{array}\right], M^{I}=\left[\begin{array}{ccc}
0 & 0& \times\\ 0 & \times & \times\\
\times& \times&\times
\end{array}\right]
\end{equation}

The Majorana Yukawa Lagrangian (LH and RH) for A1 is thus given as,

\begin{equation}\label{eqx8}
\begin{split}
\mathcal{L_{MR}}=Y_{Re\tau}{{L}_{Re}}^TCi\sigma_2\Delta_R{L}_{R\tau}+Y_{R\tau e}{{L}_{R\tau}}^TCi\sigma_2\Delta_R{L}_{Re}+ Y_{R\mu\mu}{L_{R\mu}}^TCi\sigma_2{\Delta_R}^{'}{L}_{R\mu}+\\
Y_{R\mu\tau}{{L}_{R\mu}}^TCi\sigma_2{\Delta_R}^{''}{L}_{R\tau}+Y_{R\tau\mu}{{L}_{R\tau}}^TCi\sigma_2{\Delta_R}^{''}{L}_{R\mu}+
Y_{R\tau\tau}{{L}_{R\tau}}^TCi\sigma_2{\Delta_R}^{''}{L}_{R\tau}.
\end{split}
\end{equation}

\begin{equation}\label{eqx8}
\begin{split}
\mathcal{L_{ML}}=Y_{Le\tau}{{L}_{Le}}^TCi\sigma_2\Delta_L{L}_{L\tau}+Y_{L\tau e}{{L}_{L\tau}}^TCi\sigma_2\Delta_L{L}_{Le}+ Y_{L\mu\mu}{L_{L\mu}}^TCi\sigma_2{\Delta_L}^{'}{L}_{L\mu}+\\
Y_{L\mu\tau}{{L}_{L\mu}}^TCi\sigma_2{\Delta_L}^{''}{L}_{L\tau}+Y_{L\tau\mu}{{L}_{L\tau}}^TCi\sigma_2{\Delta_L}^{''}{L}_{L\mu}+
Y_{L\tau\tau}{L_{L\tau}}^TCi\sigma_2{\Delta_L}^{'}{L}_{L\tau}.
\end{split}
\end{equation}

\textbf{Class A2:}	

For the class A2, to get the desired texture zero structures for the mass matrices, the following symmetry realization has been adopted.
\begin{table}[h!]
	\centering
	\begin{tabular}{||c| c| c| c| c| c| c| c||}
		\hline
		$l_L$ & $Z_8 \times Z_2$ & $l_R$&$Z_8 \times Z_2$ &Higgs(LH)&$Z_8 \times Z_2$&Higgs(RH)&$Z_8 \times Z_2$\\ \hline
		$l_{Le}$& $(\omega^{6},-1)$ &  $l_{Re}$&$(\omega^{2},-1)$&$\Delta_L$&$(\omega^{7},-1)$&$\Delta_R$&$(\omega,-1)$\\ \hline
		$l_{L\mu}$ &$(\omega^{3},1)$  &  $l_{R\mu}$&$(\omega^{5},1)$&${\Delta_L}^{'}$&$(\omega^{2},1)$&${\Delta_R}^{'}$&$(\omega^{6},1)$\\ \hline
		$l_{L\tau}$ & $(\omega^{3},-1)$&  $l_{R\tau}$&$(\omega^{5},-1)$&${\Delta_L}^{''} $&$ (\omega^{2},-1),   $&${\Delta_R}^{''}$&$(\omega^{6},-1)$\\ \hline

	\end{tabular}
	\caption{Particle assignments for A2} \label{t1}
\end{table}

The Majorana mass terms (LH and RH) and the type I SS mass terms, in the matrix form has been obtained as,
\begin{equation}
M_{RR}=\left[\begin{array}{ccc}
\omega^4 & 1& -1\\ 1 & 1 & 1\\
-1 & 1 & 1
\end{array}\right],M_{LL}=\left[\begin{array}{ccc}
\omega^3 & 1& -1 \\ 1 & 1 & 1\\
-1 & 1 & 1
\end{array}\right], M^{I}=\left[\begin{array}{ccc}
0 & \times& 0\\ \times & \times &\times\\
0 & \times & \times
\end{array}\right].
\end{equation}

The corresponding Majorana Yukawa Lagrangian (LH and RH) for A2 is,
\begin{equation}\label{eqx8}
\begin{split}
\mathcal{L_{MR}}=Y_{Re\mu}{{L}_{Re}}^TCi\sigma_2\Delta_R{L}_{R\mu}+Y_{R\mu e}{{L}_{R\mu}}^TCi\sigma_2\Delta_R{L}_{Re}+ Y_{R\mu\mu}{L_{R\mu}}^TCi\sigma_2{\Delta_R}^{'}{L}_{R\mu}+\\
Y_{R\mu\tau}{{L}_{R\mu}}^TCi\sigma_2{\Delta_R}^{''}{L}_{R\tau}+Y_{R\tau\mu}{{L}_{R\tau}}^TCi\sigma_2{\Delta_R}^{''}{L}_{R\mu}+
Y_{R\tau\tau}{L_{R\tau}}^TCi\sigma_2{\Delta_R}^{'}{L}_{R\tau}
\end{split}
\end{equation}

\begin{equation}\label{eqx8}
\begin{split}
\mathcal{L_{ML}}=Y_{Le\tau}{{L}_{Le}}^TCi\sigma_2\Delta_L{L}_{L\tau}+Y_{L\tau e}{{L}_{L\tau}}^TCi\sigma_2\Delta_L{L}_{Le}+ Y_{L\mu\mu}{L_{L\mu}}^TCi\sigma_2{\Delta_L}^{'}{L}_{L\mu}+\\
Y_{L\mu\tau}{{L}_{L\mu}}^TCi\sigma_2{\Delta_L}^{''}{L}_{L\tau}+Y_{L\tau\mu}{{L}_{L\tau}}^TCi\sigma_2{\Delta_L}^{''}{L}_{L\mu}+
Y_{L\tau\tau}{L_{L\tau}}^TCi\sigma_2{\Delta_L}^{'}{L}_{L\tau}
\end{split}
\end{equation}

\textbf{Class B1:}

The symmetry realizations of the particles under $Z8\times Z2$ for the class B1 are as show in the table below,
\begin{table}[h!]
	\centering
	\begin{tabular}{||c| c| c| c| c| c| c| c||}
		\hline
		$l_L$ & $Z_8 \times Z_2$ & $l_R$&$Z_8 \times Z_2$ &Higgs(LH)&$Z_8 \times Z_2$&Higgs(RH)&$Z_8 \times Z_2$\\ \hline
		$l_{Le}$&$(\omega,-1)$  &  $l_{Re}$&$(\omega^{7},-1)$&$\Delta_L$&$(\omega^{6},1)$&$\Delta_R$&$(\omega^{2},1)$\\ \hline
		$l_{L\mu}$ &$(\omega^{2},1)$  &  $l_{R\mu}$&$(\omega^{6},1)$&${\Delta_L}^{'}$&$(\omega^{5},-1)$&${\Delta_R}^{'}$&$(\omega^{3},-1)$\\ \hline
		$l_{L\tau}$ & $(\omega^{5},-1)$&  $l_{R\tau}$&$(\omega^{3},-1)$&${\Delta_L}^{''} $& $(\omega,-1)$&${\Delta_R}^{''}$&$ (\omega^{7},-1) $\\ \hline

	\end{tabular}
	\caption{Particle assignments for B1} \label{t1}
\end{table}

These transformations leads to the  Majorana mass matrices (LH and RH) as,
\begin{equation}
M_{RR}=\left[\begin{array}{ccc}
1 & 1& \omega^4\\ 1 & \omega^7 & 1\\
\omega^4 & 1 & 1
\end{array}\right],M_{LL}=\left[\begin{array}{ccc}
1 & 1& \omega^4\\ 1 & \omega & 1\\
\omega^4 & 1 & 1
\end{array}\right], M^{I}=\left[\begin{array}{ccc}
\times & \times& 0\\ \times & 0 & \times\\
0 & \times & \times
\end{array}\right]
\end{equation}
The  corresponding $Z8\times Z2$ invariant Majorana Yukawa Lagrangian (LH and RH) for B1 is,
\begin{equation}\label{eqx8}
\begin{split}
\mathcal{L_{MR}}=Y_{Ree}{{L}_{Re}}^TCi\sigma_2\Delta_R{L}_{Re}+Y_{R\tau\tau}{{L}_{R\tau}}^TCi\sigma_2\Delta_R{L}_{R\tau}+ Y_{Re\mu}{L_{Re}}^TCi\sigma_2{\Delta_R}^{'}{L}_{R\mu}+\\
Y_{R\mu e}{{L}_{R\mu}}^TCi\sigma_2{\Delta_R}^{'}{L}_{Re}+Y_{R\mu\tau}{{L}_{R\mu}}^TCi\sigma_2{\Delta_R}^{''}{L}_{R\tau}+
Y_{R\tau\mu}{L_{R\tau}}^TCi\sigma_2{\Delta_R}^{''}{L}_{R\mu}.
\end{split}
\end{equation}

\begin{equation}\label{eqx8}
\begin{split}
\mathcal{L_{ML}}=Y_{Lee}{{L}_{Le}}^TCi\sigma_2\Delta_L{L}_{Le}+Y_{L\tau\tau}{{L}_{L\tau}}^TCi\sigma_2\Delta_L{L}_{R\tau}+ Y_{Le\mu}{L_{Le}}^TCi\sigma_2{\Delta_L}^{'}{L}_{L\mu}+\\
Y_{L\mu e}{{L}_{L\mu}}^TCi\sigma_2{\Delta_L}^{'}{L}_{Le}+Y_{L\mu\tau}{{L}_{L\mu}}^TCi\sigma_2{\Delta_L}^{''}{L}_{L\tau}+
Y_{L\tau\mu}{L_{L\tau}}^TCi\sigma_2{\Delta_L}^{''}{L}_{L\mu}.
\end{split}
\end{equation}

\textbf{Class B2:}

The symmetry realizations to obtain the desired textures of the class B1 are as shown in the table below,
\begin{table}[h!]
	\centering
	\begin{tabular}{||c| c| c| c| c| c| c| c||}
		\hline
		$l_L$ & $Z_8 \times Z_2$ & $l_R$&$Z_8 \times Z_2$ &Higgs(LH)&$Z_8 \times Z_2$&Higgs(RH)&$Z_8 \times Z_2$\\ \hline
		$l_{Le}$&$(\omega^{5},-1)$  &  $l_{Re}$&$(\omega^{3},-1)$&$\Delta_L$&$(\omega^{6},1)$&$\Delta_R$&$(\omega^{2},1)$\\ \hline
		$l_{L\mu}$ &$(\omega^{3},1)$  &  $l_{R\mu}$&$(\omega^{5},1)$&${\Delta_L}^{'}$&$(\omega^{2},1)$&${\Delta_R}^{'}$&$(\omega^{6},1)$\\ \hline
		$l_{L\tau}$ & $(\omega^{5},-1)$&  $l_{R\tau}$&$(\omega^{3},-1)$&${\Delta_L}^{''} $& $(1,-1)$&${\Delta_R}^{''}$&$ (1,-1) $\\ \hline

	\end{tabular}
	\caption{Particle assignments for B2} \label{t1}
\end{table}

The Majorana mass matrices (LH and RH) and the type I SS mass matrix under these transformation has been obtained as,
\begin{equation}
M_{RR}=\left[\begin{array}{ccc}
1 & \omega^2& 1\\ \omega^2 & 1 & 1\\
1 & 1 & \omega^6
\end{array}\right],M_{LL}=\left[\begin{array}{ccc}
1 & \omega^2& 1\\ \omega^2 & 1 & 1\\
1 & 1 & \omega^2
\end{array}\right], M^{I}=\left[\begin{array}{ccc}
\times & 0& \times\\ 0 & \times & \times\\
\times & \times & 0
\end{array}\right]
\end{equation}
The corresponding Majorana Yukawa Lagrangian (LH and RH) for the class B2 is,
\begin{equation}\label{eqx8}
\begin{split}
\mathcal{L_{MR}}=Y_{Ree}{{L}_{Re}}^TCi\sigma_2\Delta_R{L}_{Re}+Y_{Re\tau}{{L}_{Re}}^TCi\sigma_2\Delta_R{L}_{R\tau}+ Y_{R\tau e}{L_{R\tau}}^TCi\sigma_2{\Delta_R}L_{Re}+\\
Y_{R\mu \mu}{{L}_{R\mu}}^TCi\sigma_2{\Delta_R}^{'}{L}_{R\mu}+Y_{R\mu\tau}{{L}_{R\mu}}^TCi\sigma_2{\Delta_R}^{''}{L}_{R\tau}+
Y_{R\tau\mu}{L_{R\tau}}^TCi\sigma_2{\Delta_R}^{''}{L}_{R\mu}
\end{split}
\end{equation}

\begin{equation}\label{eqx8}
\begin{split}
\mathcal{L_{ML}}=Y_{Lee}{{L}_{Le}}^TCi\sigma_2\Delta_L{L}_{Le}+Y_{Le\tau}{{L}_{Le}}^TCi\sigma_2\Delta_L{L}_{L\tau}+ Y_{L\tau e}{L_{L\tau}}^TCi\sigma_2{\Delta_L}{L}_{Le}+\\
Y_{L\mu \mu}{{L}_{L\mu}}^TCi\sigma_2{\Delta_L}^{'}{L}_{L\mu}+Y_{L\mu\tau}{{L}_{L\mu}}^TCi\sigma_2{\Delta_L}^{''}{L}_{L\tau}+
Y_{L\tau\mu}{L_{L\tau}}^TCi\sigma_2{\Delta_L}^{''}{L}_{L\mu}
\end{split}
\end{equation}

\textbf{Class B3:}

The symmetry realizations of the particles to obtain the desired mass terms of class B3 is shown in table V.
\begin{table}[h!]
	\centering
	\begin{tabular}{||c| c| c| c| c| c| c| c||}
		\hline
		$l_L$ & $Z_8 \times Z_2$ & $l_R$&$Z_8 \times Z_2$ &Higgs(LH)&$Z_8 \times Z_2$&Higgs(RH)&$Z_8 \times Z_2$\\ \hline
		$l_{Le}$&$(\omega^{5},-1)$  &  $l_{Re}$&$(\omega^{3},-1)$&$\Delta_L$&$(\omega^{6},1)$&$\Delta_R$&$(\omega^{2},1)$\\ \hline
		$l_{L\mu}$ &$(\omega^{3},1)$  &  $l_{R\mu}$&$(\omega^{5},1)$&${\Delta_L}^{'}$&$(1,-1)$&${\Delta_R}^{'}$&$(\omega^{6},1)$\\ \hline
		$l_{L\tau}$ & $(\omega^{3},1)$&  $l_{R\tau}$&$(\omega^{5},1)$&${\Delta_L}^{''} $& $(\omega^{2},1)$&${\Delta_R}^{''}$&$ (1,-1) $\\ \hline

	\end{tabular}
	\caption{Particle assignments for B3} \label{t1}
\end{table}

The Majorana mass terms (LH and RH) and the type I SS mass terms, in the matrix form can be written as,
\begin{equation}
M_{RR}=\left[\begin{array}{ccc}
1 & -1& 1\\ -1 & \omega^4 & 1\\
1 & 1 & 1
\end{array}\right],M_{LL}=\left[\begin{array}{ccc}
1 & \omega^6& 1\\ \omega^6 & \omega^6 & 1\\
1 & 1 & 1
\end{array}\right], M^{I}=\left[\begin{array}{ccc}
\times & 0& \times\\ 0 & 0 &\times\\
\times& \times & \times
\end{array}\right]
\end{equation}

The corresponding Majorana Yukawa Lagrangian (LH and RH) for B3 is,
\begin{equation}\label{eqx8}
\begin{split}
\mathcal{L_{MR}}=Y_{Ree}{{L}_{Re}}^TCi\sigma_2\Delta_R{L}_{Re}+Y_{Re\tau}{{L}_{Re}}^TCi\sigma_2\Delta_R^{''}{L}_{R\tau}+ Y_{R\tau e}{L_{R\tau}}^TCi\sigma_2\Delta_R^{''}{L}_{Re}+\\
Y_{R\mu \tau}{{L}_{R\mu}}^TCi\sigma_2{\Delta_R}^{'}{L}_{R\tau}+Y_{R\tau \mu}{{L}_{R\tau}}^TCi\sigma_2{\Delta_R}^{'}{L}_{R\mu}+
Y_{R\tau\tau}{L_{R\tau}}^TCi\sigma_2{\Delta_R}^{'}{L}_{R\tau}
\end{split}
\end{equation}

\begin{equation}\label{eqx8}
\begin{split}
\mathcal{L_{ML}}=Y_{Lee}{{L}_{Le}}^TCi\sigma_2\Delta_L{L}_{Le}+Y_{Le\tau}{{L}_{Le}}^TCi\sigma_2{\Delta_L}^{'}{L}_{L\tau}+ Y_{L\tau e}{L_{L\tau}}^TCi\sigma_2{\Delta_L}^{'}{L}_{Le}+\\
Y_{L\mu \tau}{{L}_{L\mu}}^TCi\sigma_2{\Delta_L}^{''}{L}_{L\tau}+Y_{L\tau \mu}{{L}_{L\tau}}^TCi\sigma_2{\Delta_L}^{''}{L}_{L\mu}+
Y_{L\tau\tau}{L_{L\tau}}^TCi\sigma_2{\Delta_L}^{''}{L}_{L\tau}
\end{split}
\end{equation}

\textbf{Class B4:}

Similarly we give the transformations for the class B4 as shown in table VI to obtain the desired texture zero mass matrices.
\begin{table}[h!]
	\centering
	\begin{tabular}{||c| c| c| c| c| c| c| c||}
		\hline
		$l_L$ & $Z_8 \times Z_2$ & $l_R$&$Z_8 \times Z_2$ &Higgs(LH)&$Z_8 \times Z_2$&Higgs(RH)&$Z_8 \times Z_2$\\ \hline
		$l_{Le}$&$(\omega^{5},-1)$  &  $l_{Re}$&$(\omega^{3},-1)$&$\Delta_L$&$(\omega^{6},1)$&$\Delta_R$&$(\omega^{2},1)$\\ \hline
		$l_{L\mu}$ &$(\omega^{3},1)$  &  $l_{R\mu}$&$(\omega^{5},1)$&${\Delta_L}^{'}$&$(\omega^{2},1)$&${\Delta_R}^{'}$&$(\omega^{6},1)$\\ \hline
		$l_{L\tau}$ & $(\omega^{5},-1)$&  $l_{R\tau}$&$(\omega^{3},-1)$&${\Delta_L}^{''} $& $(1,-1) $&${\Delta_R}^{''}$&$ (1,-1) $\\ \hline

	\end{tabular}
	\caption{Particle assignments for B4} \label{t1}
\end{table}

Under these symmetry realizations, we obtain the  Majorana mass terms (LH and RH) and the type I SS mass terms as,
\begin{equation}
M_{RR}=\left[\begin{array}{ccc}
1 & 1& \omega^4\\ 1 & 1 & 1\\
\omega^4 & 1 & \omega^6
\end{array}\right], M_{LL}=\left[\begin{array}{ccc}
1 & 1& \omega^4\\1 & 1 & 1\\
\omega^4 & 1 &\omega^2
\end{array}\right], M^{I}=\left[\begin{array}{ccc}
\times & \times& 0\\ \times& \times & \times\\
0 & \times & 0
\end{array}\right]
\end{equation}

The Majorana Yukawa Lagrangian (LH and RH) for B4 thus becomes,
\begin{equation}\label{eqx8}
\begin{split}
\mathcal{L_{MR}}=Y_{Ree}{{L}_{Re}}^TCi\sigma_2\Delta_R{L}_{Re}+Y_{Re\mu}{{L}_{Re}}^TCi\sigma_2{\Delta_R}^{''}{L}_{R\mu}+ Y_{R\mu e}{L_{R\mu}}^TCi\sigma_2{\Delta_R}^{''}{L}_{Re}+\\
Y_{R\mu \mu}{{L}_{R\mu}}^TCi\sigma_2{\Delta_R}^{'}{L}_{R\mu}+Y_{R \mu\tau}{{L}_{R\mu}}^TCi\sigma_2{\Delta_R}^{''}{L}_{R\tau}+
Y_{R\tau\mu}{L_{R\tau}}^TCi\sigma_2{\Delta_R}^{''}{L}_{R\mu}
\end{split}
\end{equation}

\begin{equation}\label{eqx8}
\begin{split}
\mathcal{L_{ML}}=Y_{Lee}{{L}_{Le}}^TCi\sigma_2\Delta_L{L}_{Le}+Y_{Le\mu}{{L}_{Le}}^TCi\sigma_2{\Delta_L}^{''}{L}_{L\mu}+ Y_{L\mu e}{L_{L\mu}}^TCi\sigma_2{\Delta_L}^{''}{L}_{Le}+\\
Y_{L\mu \mu}{{L}_{L\mu}}^TCi\sigma_2{\Delta_L}^{'}{L}_{L\mu}+Y_{L \mu\tau}{{L}_{L\mu}}^TCi\sigma_2{\Delta_L}^{''}{L}_{L\tau}+
Y_{L\tau\mu}{L_{L\tau}}^TCi\sigma_2{\Delta_L}^{''}{L}_{L\mu}
\end{split}
\end{equation} 	

The type II SS mass term in LRSM is directly proportional to the Majorana mass term as evident from equation \ref{eqx14}, so it will have the same structure as $M_{RR}$ and $M_{LL}$. LRSM being a combination of type I and type II SS mass terms would give us the final mass matrix that would obey the structure of two zero texture mass matrix. It has been shown in tabular form in the numerical analysis.

\section{NEUTRINOLESS DOUBLE BETA DECAY AND LEPTON FLAVOUR VIOLATION IN LEFT-RIGHT SYMMETRIC MODEL}{\label{sec:level5}}

The very facts of LRSM and the presence of several new heavy particles leads to many new contributions to NDBD apart from the standard light neutrino contribution. This has been extensively studied in several earlier works \cite{barry2013lepton,borah2015neutrinoless,awasthi2016implications,chakrabortty2012neutrinoless,borah2015neutrinoless,patra2013neutrinoless,parida2013left,bambhaniya2016scalar,borgohain2017neutrinoless,borgohain2017lepton}. Amongst the non standard contribution, notable are,  heavy RH  neutrino contribution to NDBD in which the mediator particles are the $\rm {W_L}^-$ and $\rm {W_R}^-$ boson individually, light neutrino contribution to NDBD in which the intermediate particles are $\rm {W_R}^-$ bosons, light neutrino contribution  mediated by both $ \rm {W_L}^-$ and  $\rm {W_R}^-$, heavy neutrino contribution  mediated by both  $ \rm {W_L}^-$ and  $\rm {W_R}^-$,  triplet Higgs $\rm \Delta_L$ contribution mediated by $ \rm{W_L}^-$  bosons and RH triplet Higgs $\rm \Delta_R$ contribution to NDBD in which the mediator particles are $ {W_R}^-$  bosons .
The amplitude of
these processes are dependent on the mixing between light and heavy neutrinos, the mass of the heavy neutrino, $\rm N_i$, the mass of the gauge bosons, $\rm {W_L}^-$ and $\rm {W_R}^-$,
 the elements of the RH leptonic mixing matrix,
LH and RH triplet Higgs, $\rm \Delta_L$ and $\rm \Delta_R$  as well as their coupling to leptons, $\rm f_L$ and  $f_R$.

\par Besides the observation of neutrino oscillation also provides compelling evidence for charged lepton flavour violation (CLFV) \cite{cirigliano2004lepton,calibbi2017charged,bernstein2013charged}. Since LFV which is generated at high energy scales are beyond the reach of the colliders, searching them in low energy scales amongst the charged leptons is widely accepted as an alternate procedure to probe LFV at high scales. Many previous works \cite {borgohain2017neutrinoless,borgohain2017lepton,barry2013lepton,bambhaniya2016scalar,perez2017lepton,deppisch2015double}have focussed on the lepton flavour violating decay modes of muon, ( $\mu\rightarrow 3e $ , $\mu\rightarrow e\gamma $, $\mu\rightarrow e$ conversion in the nuclei). Considerable CLFV occurs in LRSM owing to the contributions that arises from the heavy RH neutrino and Higgs scalars. The relevant branching ratios (BR) has been derived and studied in \cite{cirigliano2004lepton}. The LFV processes $\mu\rightarrow 3e $, $\mu\rightarrow e\gamma $ provides the most relevant constraints on the masses of the RH neutrinos and the doubly charged scalars. In this work we would consider the process $\mu\rightarrow e\gamma $, the BR of which is given by,

\begin{equation}\label{eqa}
BR\left(\mu\rightarrow e\gamma\right)= 1.5\times 10^{-7}{\left|g_{lfv}\right|}^2{\left(\frac{1 TeV}{M_{W_R}}\right)}^4,
\end{equation}\\
where, $ g_{lfv}$ is defined as,
\begin{equation}\label{eqb}
g_{lfv}=\sum_{n=1}^{3}V_{\mu n}{V_{e n}}^*{\left(\frac{M_n}{M_{W_R}}\right)}^2=\frac{\left[M_R {M_R}^*\right]_{\mu e}}{{M_{W_R}}^2}.
\end{equation}

The current experimental constraints for the BRs of these processes has been obtained as  $<1.0\times 10^{-12}$ for $\mu\rightarrow 3e $ at $ 90\%$ CL was obtained  by the
SINDRUM experiment. While it is $<4.2\times 10^{-13}$  \cite{baldini2016search} for the process $\mu\rightarrow e\gamma$, established by the MEG collaboration.

	\section{NUMERICAL ANALYSIS AND RESULTS}

\begin{itemize}	
	\item  In LRSM, we can write the light neutrino mass matrix as a combination of type I and type II mass terms as,
	\begin{equation}\label{a1}
	\rm M_{\nu}={M_{\nu}}^I+{M_{\nu}}^{II}
	\end{equation}
	Here, we consider $\rm {M_{\nu}}^I$ to be favoring the TM mixing with magic symmetry, so as to obtain the desired two zero texture. The different magic neutrino mass matrix with two zeroes can be obtained from the most general magic mass matrix which can be parameterized as\cite{gautam2018trimaximal,gautam2016zeros},
	\begin{equation}
	\rm M_{magic}=\left[\begin{array}{ccc}
	p & q & r\\
	q& r& p+r-s\\
	r & p+r-s& q-r+s
	\end{array}\right]
	\end{equation}
	which can be diagonalized by the trimaximal mixing matrix as,
	
	$\rm M_{diag}={{U_{TM}}_2}^T M_{magic} {U_{TM}}_2$ where, $\rm {U_{TM}}_2$ is the diagonalizing matrix for the magic mass matrix and is given in equation \ref{y1}
		\item Using the constraint relations for various classes with two zero textures, we can arrive at the mass matrices as,
	
	\begin{equation}
\rm	{M_{\nu}}^I(A1)=\left[\begin{array}{ccc}
	0 & 0 & r\\
	0& s& r-s\\
	r & r-s& -r+s
	\end{array}\right], {M_{\nu}}^I(A2)=\left[\begin{array}{ccc}
	0 & q & 0\\
	q& s& -s\\
	0 & -s& q+s
	\end{array}\right]
	\end{equation}
	
	\begin{equation}
\rm	{M_{\nu}}^I(B1)=\left[\begin{array}{ccc}
	p & q & 0\\
	q& 0& p\\
	0 & p& q
	\end{array}\right], {M_{\nu}}^I(B2)=\left[\begin{array}{ccc}
	p & 0 & r\\
	0& r& p\\
	r & p& 0
	\end{array}\right]
	\end{equation}

	\begin{equation}
\rm	{M_{\nu}}^I(B3)=\left[\begin{array}{ccc}
	p & 0 & r\\
	0& 0& p+r\\
	r & p+r& -r
	\end{array}\right], {M_{\nu}}^I(B4)=\left[\begin{array}{ccc}
	p & q & 0\\
	q& -q& p+q\\
	0 & p+q& 0
	\end{array}\right]
	\end{equation}
	
	Again , $\rm {M_{\nu}}^I={U_{TM}}_2U_{Maj}{M_{\nu}}^{diag}{U_{Maj}}^T {U_{TM}}_2^T$ where,$U_{Maj}$ consists of the Majorana phases $\alpha$ and $\beta$, $\rm {M_{\nu}}^{diag}= $diag $(m_1, m_2, m_3)$ which can be written as,
	\begin{itemize}
		\item diag $(m_1, \sqrt{m_1^2+\Delta m_{sol}^2},\sqrt{m_1^2+\Delta m_{sol}^2+\Delta m_{atm}^2} )$ (In NH),
		\item diag $(\sqrt{m_3^2+\Delta m_{atm}^2}, \sqrt{m_3^2+\Delta m_{sol}^2+\Delta m_{atm}^2} ,m_3 )$ (In IH),
	\end{itemize}

	in terms of the lightest neutrino mass. Thus, by comparing ${M_{\nu}}^I$ with ${M_{\nu}}^I$ for different classes  we can solve for the unknown parameters (p, q, r, s) in the corresponding matrices and obtain $\rm {M_{\nu}}^I$ for different classes.
	
	\item Since now we have $\rm {M_{\nu}}^I$, we can evaluate 	$\rm {M_{\nu}}^{II}$ using equation \ref{a1}. Again, we have in LRSM, $\rm M_{RR}=\gamma(\frac{M_{W_R}}{M_{W_L}})^2 {M_{\nu}}^{II}$,
	where $\gamma$ is a dimensionless parameter which follows directly from the minimization of the  Higgs potential, here we consider its value to be $10^{-10}$. Thus we can find out $\rm M_{RR}$
	for our further analysis.
\item Using the constraint relations  for the respective classes, we have compared the neutrino mass matrix, $\rm M_\nu=U_{PMNS}{M_{\nu}}^{diag}{U_{PMNS}}^T$ with the neutrino mass matrices ($\rm {M_{\nu}}^I+{M_{\nu}}^{II}$) containing two zeros. $ \rm U_{PMNS}$ being the diagonalizing matrix of the light neutrino mass matrix, $M_\nu$ and is given by,

\begin{equation}\label{eq5}
\rm U_{PMNS}=\left[\begin{array}{ccc}
c_{12}c_{13}&s_{12}c_{13}&s_{13}e^{-i\delta}\\
-c_{23}s_{12}-s_{23}s_{13}c_{12}e^{i\delta}&-c_{23}c_{12}-s_{23}s_{13}s_{12}e^{i\delta}&s_{23}c_{13}\\
s_{23}s_{12}-c_{23}s_{13}c_{12}e^{i\delta}&-s_{23}c_{12}-c_{23}s_{13}s_{12}e^{i\delta}&c_{23}c_{13}
\end{array}\right]U_{Maj}.
\end{equation}

The abbreviations used here are $c_{ij}$= $\cos\theta_{ij}$, $s_{ij}$=$\sin\theta_{ij}$, $\delta$ is the Dirac CP phase.
$ \rm U_{Maj}$ is $ \rm diag (1,e^{i\alpha},e^{i\beta}) $
contains the Majorana phases $ \rm \alpha$ and $ \rm \beta$. Varying the parameters, $\theta_{12}$, $\theta_{13}$, $\delta$ in its 3$\sigma$ range \cite{de2018status} and writing the mass Eigen values in terms of lightest neutrino mass $m_1/m_3$ for (NH/IH) and varying from 0.0001 to 0.1, we have solved for the parameters  $ \rm \alpha$, $\beta$ and $\theta_{23}$. We have chosen these parameters as the Majorana phases are unknown yet and the precise measurement of $\theta_{23}$ and octant degeneracy is yet to be determined although experiments like NOvA, T2K have reported some values.
	
\item The different structures of the neutrino mass matrix in the LRSM using two texture zero are shown in table \ref{t5}. The symmetry realizations of the texture zeros using the cyclic groups $Z8\times Z2$ are as shown in the previous section .

	\begin{table}[h!]
		\centering
		\begin{tabular}{| c|c| c| c| c| c| c||}
			\hline
			Class&$M_D$ & $M_{RR}$ & ${M_{\nu}}^{I}$&${M_{\nu}}^{II}$ &$M_{\nu}$\\ \hline
			A1&	$\left[\begin{array}{ccc}
			x & 0 & 0\\
			0& y& 0\\
			0 & 0& z
			\end{array}\right]$&$\left[\begin{array}{ccc}
			0 & 0 & A\\
			0& B& C\\
			A & C& D
			\end{array}\right]$ & $\left[\begin{array}{ccc}
			0 & 0 & a\\
			0& b& c\\
			a & c& d
			\end{array}\right]$&$\left[\begin{array}{ccc}
			0 & 0 & W\\
			0& X& Y\\
			W & Y& Z
			\end{array}\right]$&$\left[\begin{array}{ccc}
			0 & 0 & W+a\\
			0& X+b& Y+c\\
			W+a & Y+c& Z+d
			\end{array}\right]$\\ \hline
			A2	&	$\left[\begin{array}{ccc}
			x & 0 & 0\\
			0& y& 0\\
			0 & 0& z
			\end{array}\right]$&$\left[\begin{array}{ccc}
			0 & A & 0\\
			A& B& C\\
			0 & C& D
			\end{array}\right]$ & $\left[\begin{array}{ccc}
			0 & a & 0\\
			a& b& c\\
			0 & c& d
			\end{array}\right]$&$\left[\begin{array}{ccc}
			0 & W & 0\\
			W& X& Y\\
			0 & Y& Z
			\end{array}\right]$&$\left[\begin{array}{ccc}
			0 & W+a & 0\\
			W+a& X+b& Y+c\\
			0 & Y+c& Z+d
			\end{array}\right]$\\ \hline
			B1	&	$\left[\begin{array}{ccc}
			x & 0 & 0\\
			0& y& 0\\
			0 & 0& z
			\end{array}\right]$&$\left[\begin{array}{ccc}
			A & B & 0\\
			B& 0& C\\
			0 & C& D
			\end{array}\right]$ & $\left[\begin{array}{ccc}
			a & b & 0\\
			b& 0& c\\
			0 & c& d
			\end{array}\right]$&$\left[\begin{array}{ccc}
			W & X & 0\\
			X& 0& Y\\
			0 & Y& Z
			\end{array}\right]$&$\left[\begin{array}{ccc}
			W+a & X+b & 0\\
			X+b& 0& Y+c\\
			0 & Y+c& Z+d
			\end{array}\right]$\\ \hline
			B2&	$\left[\begin{array}{ccc}
			x & 0 & 0\\
			0& y& 0\\
			0 & 0& z
			\end{array}\right]$&$\left[\begin{array}{ccc}
			A & 0 & B\\
			0& C& D\\
			B & D& 0
			\end{array}\right]$ & $\left[\begin{array}{ccc}
			a & 0 & b\\
			0& c& d\\
			b & d& 0
			\end{array}\right]$&$\left[\begin{array}{ccc}
			W & 0 & X\\
			0& Y& Z\\
			X & Z& 0
			\end{array}\right]$&$\left[\begin{array}{ccc}
			W+a & 0 & X+b\\
			0& Y+c& Z+d\\
			X+b & Z+d& 0
			\end{array}\right]$\\ \hline
			B3	&	$\left[\begin{array}{ccc}
			x & 0 & 0\\
			0& y& 0\\
			0 & 0& z
			\end{array}\right]$&$\left[\begin{array}{ccc}
			A & 0 & B\\
			0& 0& C\\
			B & C& D
			\end{array}\right]$ & $\left[\begin{array}{ccc}
			a & 0 & b\\
			0& 0& c\\
			b & c& d
			\end{array}\right]$&$\left[\begin{array}{ccc}
			W & 0 & X\\
			0& 0& Y\\
			X & Y& Z
			\end{array}\right]$&$\left[\begin{array}{ccc}
			W+a & 0 & X+b\\
			0& 0& Y+c\\
			X+b & Y+c& Z+d
			\end{array}\right]$\\ \hline
			B4	&	$\left[\begin{array}{ccc}
			x & 0 & 0\\
			0& y& 0\\
			0 & 0& z
			\end{array}\right]$&$\left[\begin{array}{ccc}
			A & B & 0\\
			B& C&D\\
			0& D& 0
			\end{array}\right]$ & $\left[\begin{array}{ccc}
			a & b & 0\\
			b& c& d\\
			0 & d& 0
			\end{array}\right]$&$\left[\begin{array}{ccc}
			W & X & 0\\
			X& Y& Z\\
			0 & Z& 0
			\end{array}\right]$&$\left[\begin{array}{ccc}
			W+a & X+b & 0\\
			X+b& Y+c& Z+d\\
			0 & Z+d& 0
			\end{array}\right]$\\ \hline

	    \end{tabular}
		\caption{The structures of $\rm M_D$, $\rm M_{RR}$, $\rm {M_\nu}^{I}$ and $\rm {M_\nu}^{II}$  and $\rm M_\nu$ for different classes of two zero textures.} \label{t5}
	\end{table}

\item Owing to the presence of new scalars and gauge bosons in the LRSM, various additional sources would
give rise to contributions to NDBD process, which involves RH neutrinos, RH gauge bosons, scalar Higgs triplets as well as the mixed LH-RH contributions. We will study
LNV (NDBD) for the non standard contributions for the effective mass  in
the framework of LRSM. For a simplified
analysis we would ignore the left-right gauge boson mixing ($W_L-W_R$) which is very less and heavy light neutrino mixing which is dependent upon $\rm \frac{M_D}{M_R}$ is $\zeta\approx 10^{-6}$.
Furthermore, contributions from the left handed Higgs triplets is suppressed by the light neutrino mass. Thus considering the mixing between LH and RH sector to be so small, their contributions to $0\nu\beta\beta$ can be neglected. The total effective mass is thus given by the formula as used in earlier works like, \cite{borah2015neutrinoless,borgohain2017neutrinoless}
\begin{equation}\label{a2}
\rm {m_{N+\Delta}}^{eff}=p^2\frac{{M_{W_L}}^4}{{M_{W_R}}^4}\frac{{U_{Rei}}^*2}{M_i}+p^2\frac{{M_{W_L}}^4}{{M_{W_R}}^4}{{U_{Rei}}^2}M_i\left(\frac{1}{{M_{\Delta_R}}^2}+\frac{1}{{M_{{\Delta_R}^{'}}}^2}+\frac{1}{{M_{{\Delta_R}^{''}}}^2}\right).
\end{equation}
\par \hspace{6mm}Here, $\Delta$ in LHS represents the three RH scalar triplets $\rm \Delta_R$, $\rm {\Delta_R}^{'}$ and $\rm {\Delta_R}^{''}$,  $\rm <p^2> =m_e m_p \frac{M_N}{M_\nu}$ is the typical momentum exchange of the process, where $\rm m_p$ and $\rm m_e$ are the mass of the proton and electron respectively
and $ \rm M_N$ is the NME corresponding to the RH neutrino exchange. $\rm  U_{Rei}$ in equation (\ref{a2}) denotes the elements of the first row of the unitary matrix
diagonalizing the right handed neutrino mass matrix $\rm M_{RR}$ with mass Eigen values $ \rm M_i$. Since we have $\rm M_{RR}$,
we can evaluate $\rm U_{Rei}$ by diagonalizing it as, $\rm M_{RR}=U_R {M_{RR}}^{diag}{U_R}^T$.
The $\rm M_{RR}$ we obtain would consists of the mixing angles in the TM mass matrix, $\theta$ and $\phi$ along with the other parameters of our concern. As shown in paper \cite{gautam2018trimaximal,grimus2008model}, $\theta$ and $\phi$ are related to the oscillation parameters $\theta_{23}$ and $\theta_{12}$ as,
	
	\begin{equation}
\rm {Sin \theta_{12}}^2=\frac{1}{3-2 Sin^2\theta}, \rm {Sin \theta_{23}}^2=\frac{1}{2}(1+\frac{\sqrt{3}Sin 2 \theta Cos \phi}{3-2 Sin^2 \theta})
\end{equation}
We thus obtained the parameter space  for $\theta$ and $\phi$ by varying the  parameters $\theta_{12}$ and $\theta_{23}$ in its 3 $\sigma$ range which is shown in figure 1. We have seen that the trimaximal mixing angle $\theta$ lies within the range (0.05 to 0.5) radian for the 3$\sigma$ range of the solar mixing angle $\theta_{12}$ although it doesn't show significant dependence. The other mixing angle $\phi$ shows some dependence on the atmospheric mixing angle $\theta_{23}$ for both normal and inverted ordering of neutrino mass. It's values lies within (1.56-1.66)radian for the 3$\sigma$ range of  $\theta_{23}$. The plot shows an exponential decrease and then increase in $\phi$ with the increase in $\theta_{23}$ with a fall at around the best fit value.

  The effective mass governing NDBD from the new physics contribution coming from RH neutrino and scalar triplet can be obtained from equation \ref{a2}. We have shown the two parameter contour plots with effective Majorana mass as the contour as in figures 2 to 19. In figures 2 to 7, we have shown the two parameter space for $m_{lightest}$ Vs $\phi$, $\beta$ Vs $\theta$ and $\alpha$ Vs $\beta$ for both the mass hierarchies for different classes of allowed two texture zero neutrino mass. The KamLAND-Zen upper limit for the effective mass is shown in the contour.

 \item In figure 2 and 3, it is seen that the value of lightest neutrino mass ranging from (0.01 to 0.1) eV satisfies the KamLAND-Zen limit of effective mass in all the classes irrespective of the mass hierarchies. Whereas, the TM mixing angle $\phi$ for all the classes  shows  different results. In NH, for the classes A1, B1 and B4, the range of $\phi$ satisfying the experimental bound of effective mass lies from around (1.57-1.65) radian and for the classes A2, B2 and B3, it is around (1.62-1.66)radian. For IH again the classes A1, B1 and B2 has $\phi$  around (1.57-1.6) radian whereas A2, B3 and B4 has the range (1.62-1.66) for $\phi$ satisfying KamlAND-Zen limit.

 \item\par In figure 4 and 5, i.e,  $\beta$ Vs $\theta$ plot, it is seen that for NH, the classes A1, A2, B1, B2, B3 has $\theta$ ranging from (0.05-0.35) radian whereas B4 has $\theta$ from (0.35-0.55) radian which satisfies the experimental bounds of effective neutrino mass. For IH, A1, A2, B1, B2 and B3 has $\theta$ from (0.05-0.3) radian whereas B4 has the range (0.45-0.55) radian. Similarly the value of the Majorana phase $\beta$ is also constrained as seen from these plots. It is around (1-3)radian for classes A1, A2, B1, B2 and B3 for NH and (0.7-3) radian for the class B4 which satisfies the KamLAND-Zen bound. For IH again, A1, B1, B2 and B3 has range (1-3) radian, for A2 and B4 it is (0.5-3) radian and (0.7-3) radian respectively. 
 \par
\item In figure 6 and 7, again we see that the value of the Majorana phase $\alpha$ is also constrained for the experimentally allowed range of effective mass. It is different for the different classes of allowed two zero texture neutrino mass. We have summarized the range of the parameters satisfying the experimental bound of effective neutrino mass governing NDBD in table 9 and 10

 \par
 \item Figures 8 to 19 shows the two parameter contour plots with the new physics contribution to effective mass as the contour, where (0.061-0.1) eV is the KamLAND-Zen upper limit for effective neutrino mass governing NDBD. The parameters shown being the model parameters that appears in the type II SS mass matrix as shown in table 7. Since there are four parameters, W, X, Y, Z in the type II SS mass matrix, there would be $^4C_2$, i.e., 6 combinations of two parameters, which we have shown in these plots. The figures corresponds to normal and inverted hierarchies which we have shown using different contours to distinguish them. The values of these parameters which gives effective mass within  experimental bounds is summarized in table 8. Although all the classes (A1, A2, B1, B2, B3, B4) gives the allowed values of effective mass, in some cases the values are so much constrained like B1 (IH), especially in the W-X plot. Also, in the case of B2 (NH), W-X, W-Z and X-Z plots has extremely constrained parameter space.

 \item For lepton flavour violation , we have evaluated the BR for the process  $\mu\rightarrow e\gamma$ using equation \ref{eqa},
  Where V is the mixing matrix
 of the right handed neutrinos with the electrons and muons. $ \rm M_n(n=1,2,3) $ are the right handed neutrino masses. We evaluated the BR with the Majorana phase $\beta$ and atmospheric mixing angle $\theta_{23}$.
 Figures 20 and 21 shows the contour plot with BR for the decay process $(\mu\rightarrow e\gamma)$ as the contour where
 	$4.2\times 10^{-13}$ is the upper limit of BR as given by MEG experiment. After analyzing all the LFV plots for all the classes of two zero texture neutrino mass matrix, it is interestingly seen that most of the classes are unable to give BR within the limit propounded by experiment. The plots clearly excludes A1, B1, B3, B4 for IH and A2 and B4 for NH in explaining LFV as far as experimental bounds are concerned, we are considering the bound given by the MEG experiment. Out of all the classes, only the class B2 for both the hierarchies results in more parameter space satisfying the experimental bound of BR. On careful observation of the figures, we see that for the 3$\sigma$ range of $\theta_{23}$, the value of the Majorana phase, $
 	\beta$ is constrained to 1 to 3 radian.
\end{itemize}

	\begin{figure}[h!]
		\centering
		\includegraphics[width=0.4\textwidth,height=4cm]{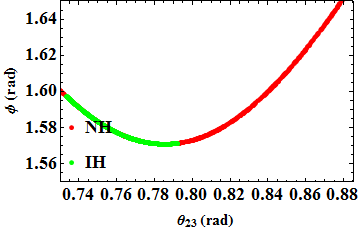}
		\includegraphics[width=0.4\textwidth,height=4cm]{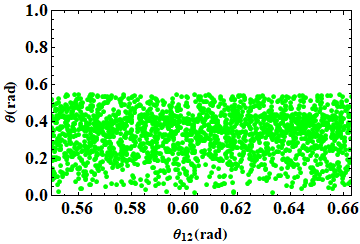}

		\caption{Variation of the TM mixing angle $\phi$ and $\theta$ with the atmospheric and solar mixing angle, $\theta_{23}$ and $\theta_{12}$.} \label{fig1}
	\end{figure}
\clearpage
	\begin{figure}[h!]
		\includegraphics[width=0.32\textwidth,height=4cm]{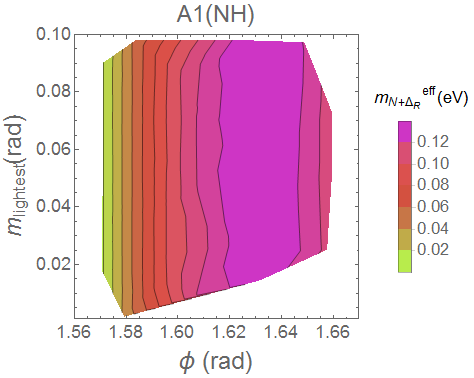}
	\includegraphics[width=0.32\textwidth,height=4cm]{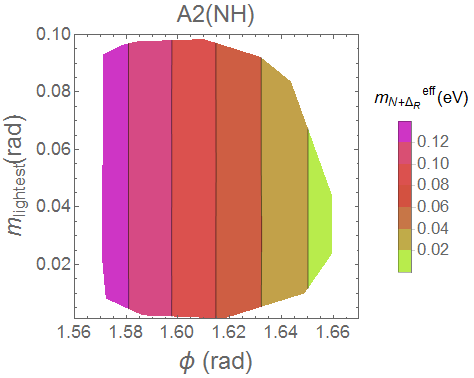}
	\includegraphics[width=0.32\textwidth,height=4cm]{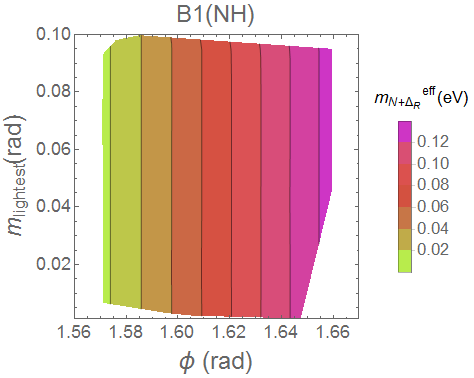}\\
	\includegraphics[width=0.32\textwidth,height=4cm]{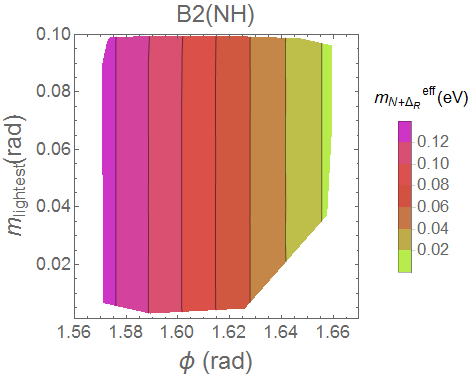}
	\includegraphics[width=0.32\textwidth,height=4cm]{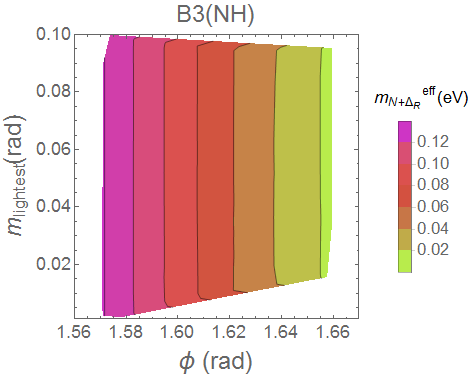}
	\includegraphics[width=0.32\textwidth,height=4cm]{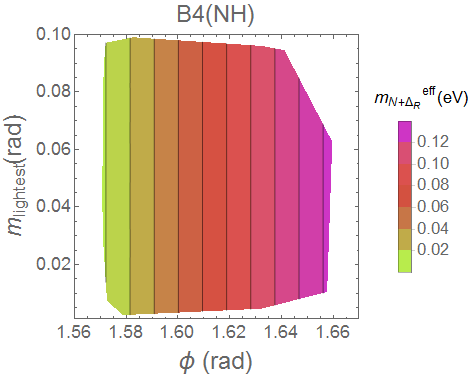}

	\caption{New physics contribution to effective mass governing NDBD for different classes of two zero textures for NH shown as a function of two  parameter $m_{lightest} $ and $ \phi$. The contour represents the  effective Majorana mass where 0.061 eV is the KamLAND-Zen upper limit.} \label{figa}
	\end{figure}

	\begin{figure}[h!]
	\includegraphics[width=0.32\textwidth,height=4cm]{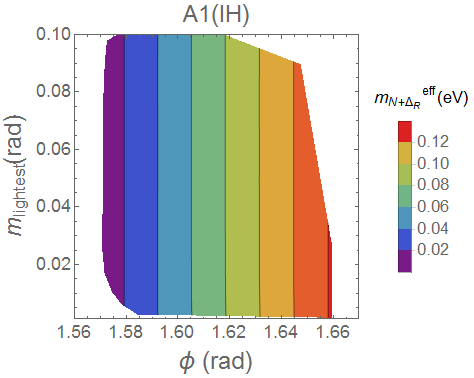}
	\includegraphics[width=0.32\textwidth,height=4cm]{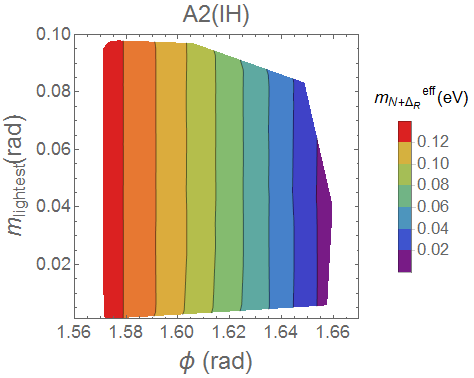}
	\includegraphics[width=0.32\textwidth,height=4cm]{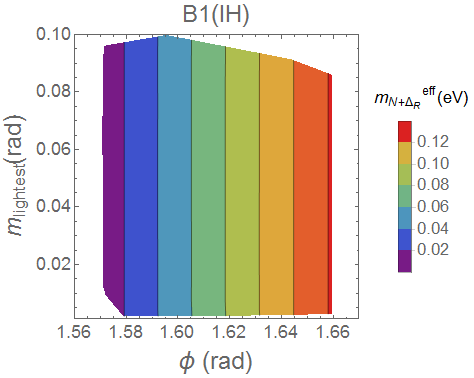}\\
	\includegraphics[width=0.32\textwidth,height=4cm]{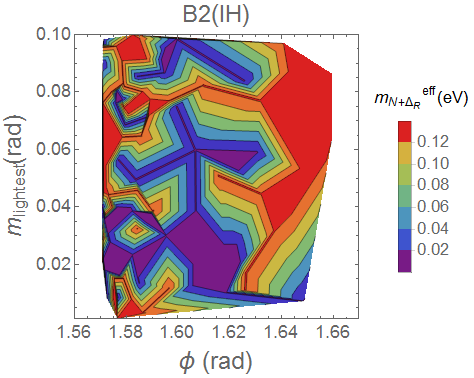}
	\includegraphics[width=0.32\textwidth,height=4cm]{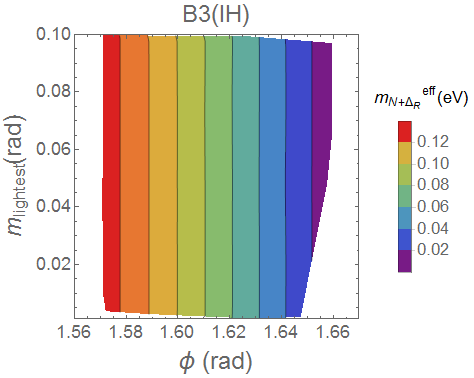}
	\includegraphics[width=0.32\textwidth,height=4cm]{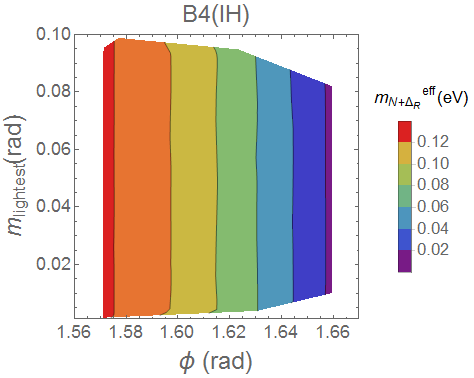}

	\caption{New physics contribution to effective mass governing NDBD for different classes of two zero textures for IH shown as a function of two  parameter $m_{lightest} $ and $ \phi$. The contour represents the  effective Majorana mass where 0.061 eV is the KamLAND-Zen upper limit.} \label{fig2}
\end{figure}
	\clearpage	
	\begin{figure}[h!]
		\includegraphics[width=0.32\textwidth,height=4cm]{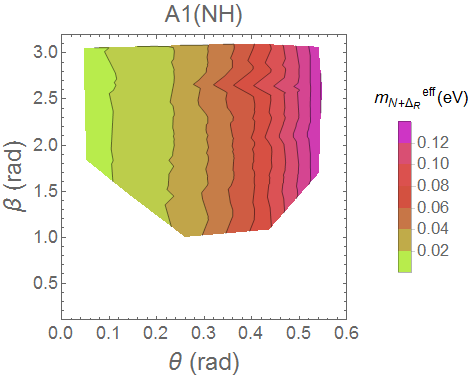}
		\includegraphics[width=0.32\textwidth,height=4cm]{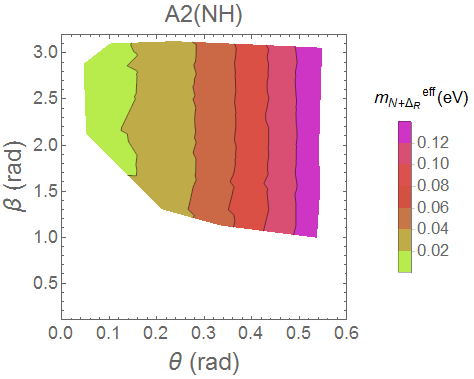}
		\includegraphics[width=0.32\textwidth,height=4cm]{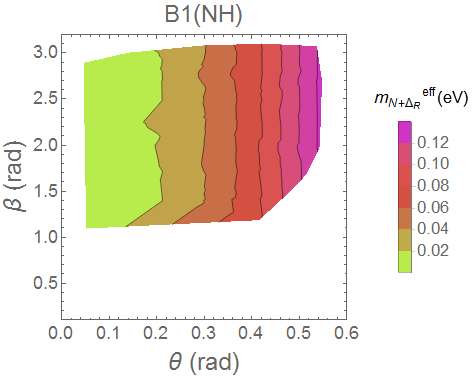}\\
		\includegraphics[width=0.32\textwidth,height=4cm]{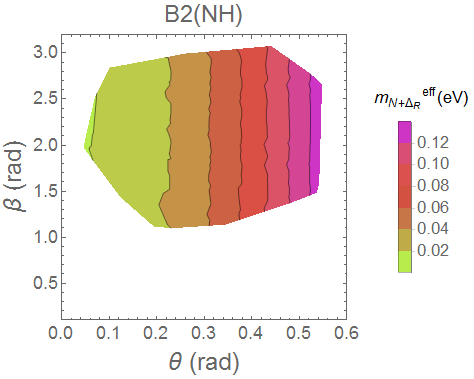}
		\includegraphics[width=0.32\textwidth,height=4cm]{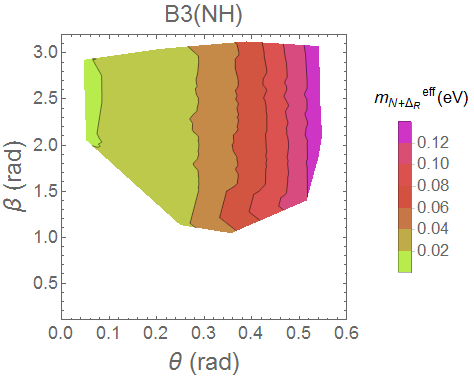}
		\includegraphics[width=0.32\textwidth,height=4cm]{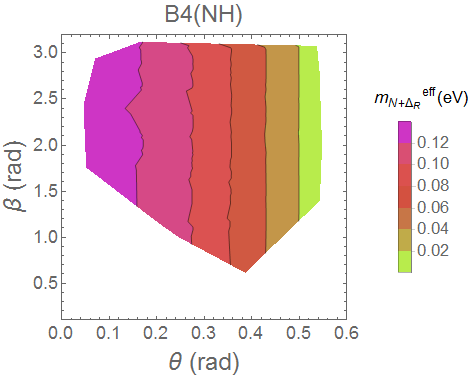}

		\caption{New physics contribution to effective mass governing NDBD for different classes of two zero textures for NH shown as a function of two  parameter $\theta $ and $ \beta$. The contour represents the  effective Majorana mass where 0.061 eV is the KamLAND-Zen upper limit.} \label{fig3}
	\end{figure}	
	
	\begin{figure}[h!]
		\includegraphics[width=0.32\textwidth,height=4cm]{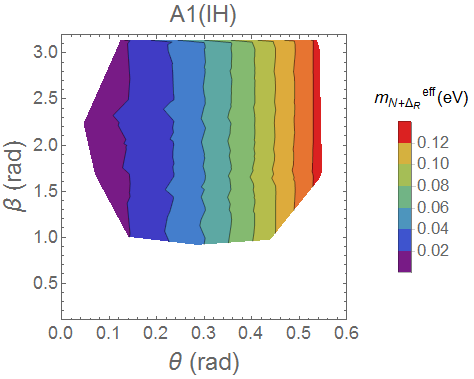}
		\includegraphics[width=0.32\textwidth,height=4cm]{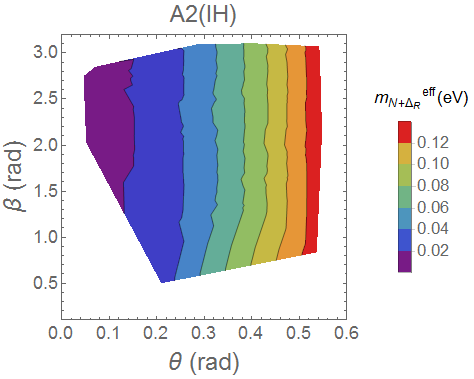}
		\includegraphics[width=0.32\textwidth,height=4cm]{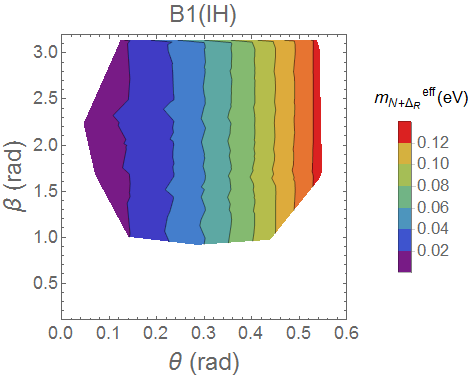}\\
		\includegraphics[width=0.32\textwidth,height=4cm]{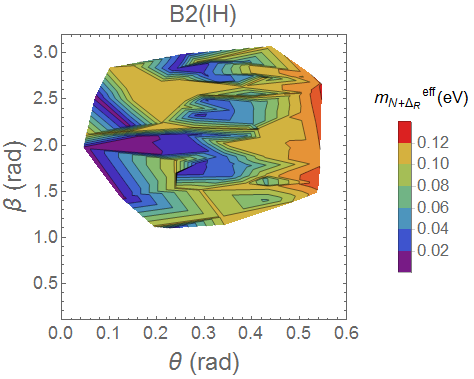}
		\includegraphics[width=0.32\textwidth,height=4cm]{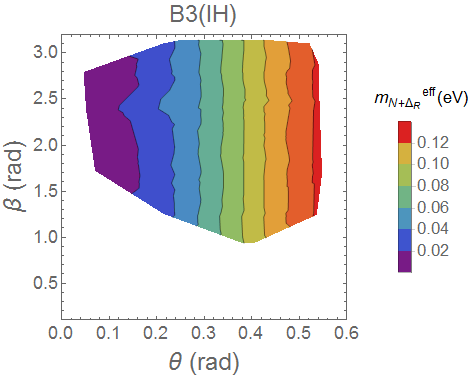}
		\includegraphics[width=0.32\textwidth,height=4cm]{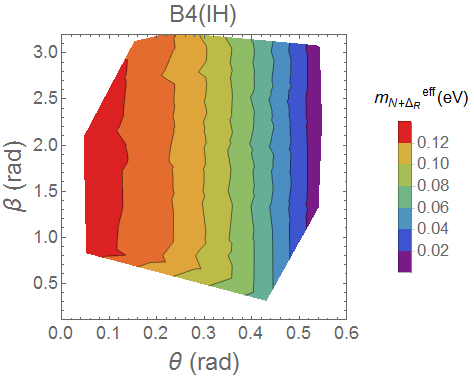}

		\caption{New physics contribution to effective mass governing NDBD for different classes of two zero textures for IH shown as a function of two  parameter $\theta $ and $ \beta$. The contour represents the  effective Majorana mass where 0.061 eV is the KamLAND-Zen upper limit.} \label{fig4}
	\end{figure}	
	\clearpage
		\begin{figure}[h!]
		\includegraphics[width=0.32\textwidth,height=4cm]{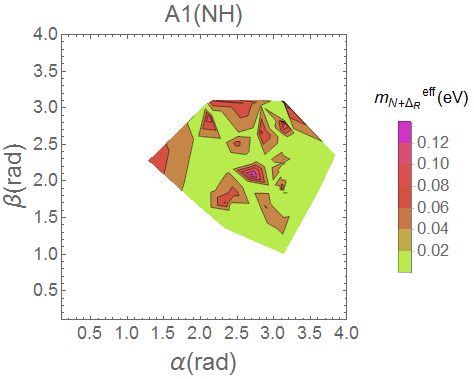}
		\includegraphics[width=0.32\textwidth,height=4cm]{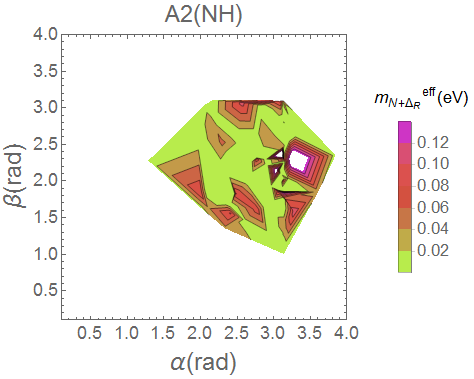}
		\includegraphics[width=0.32\textwidth,height=4cm]{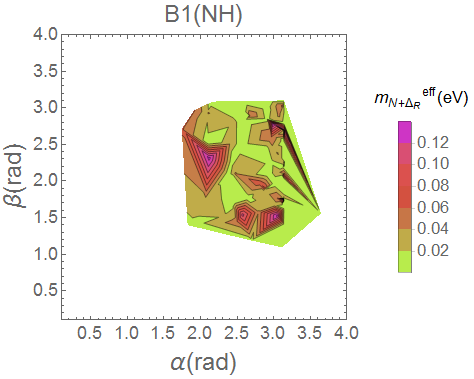}\\
		\includegraphics[width=0.32\textwidth,height=4cm]{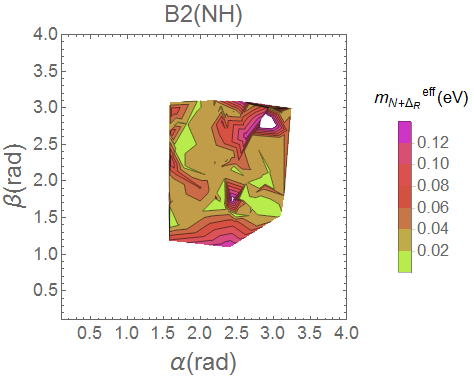}
		\includegraphics[width=0.32\textwidth,height=4cm]{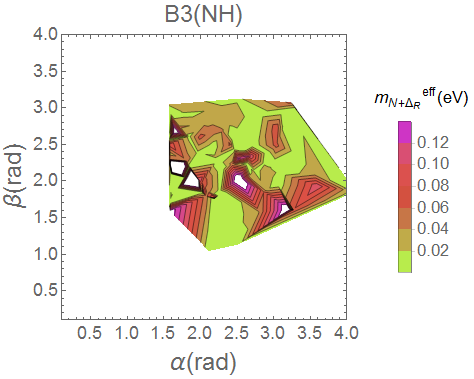}
		\includegraphics[width=0.32\textwidth,height=4cm]{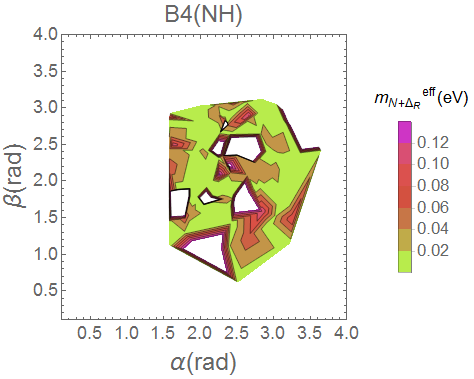}

		\caption{New physics contribution to effective mass governing NDBD for different classes of two zero textures for NH shown as a function of two  parameter $\alpha $ and$ \beta$. The contour represents the  effective Majorana mass where 0.061 eV is the KamLAND-Zen upper limit.} \label{fig4}
	\end{figure}	
	
		\begin{figure}[h!]
		\includegraphics[width=0.32\textwidth,height=4cm]{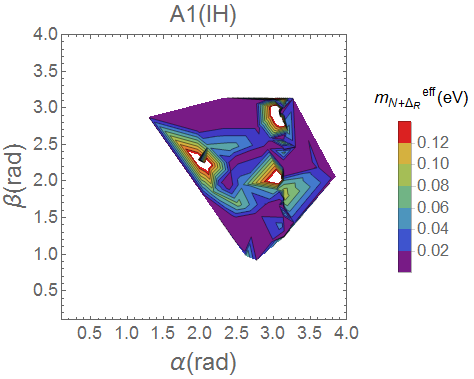}
		\includegraphics[width=0.32\textwidth,height=4cm]{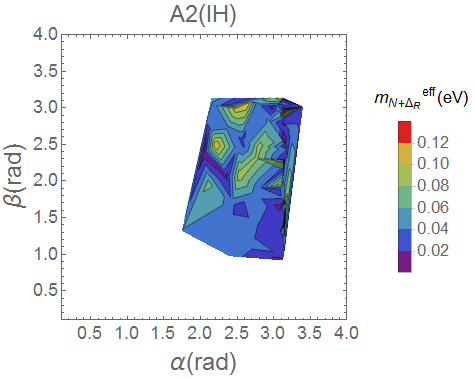}
		\includegraphics[width=0.32\textwidth,height=4cm]{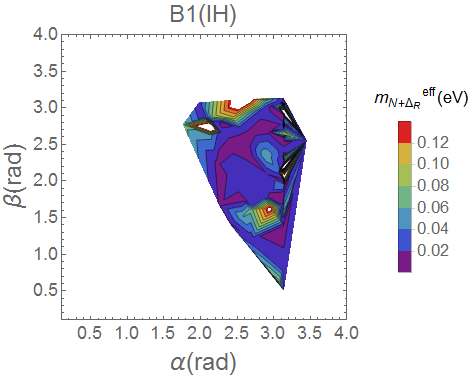}\\
		\includegraphics[width=0.32\textwidth,height=4cm]{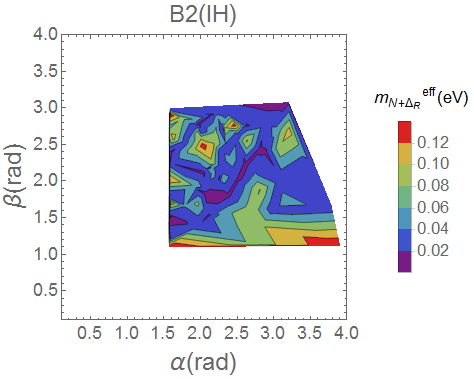}
		\includegraphics[width=0.32\textwidth,height=4cm]{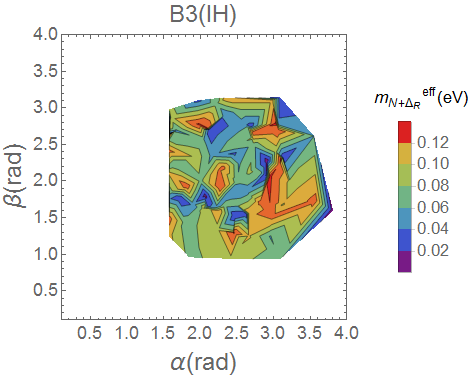}
		\includegraphics[width=0.32\textwidth,height=4cm]{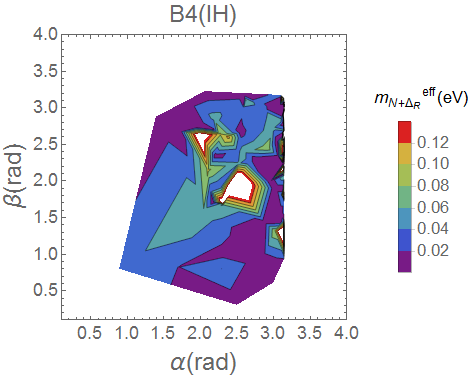}

		\caption{New physics contribution to effective mass governing NDBD for different classes of two zero textures for IH shown as a function of two  parameter $\alpha $ and$ \beta$. The contour represents the  effective Majorana mass where 0.061 eV is the KamLAND-Zen upper limit.} \label{fig4}
	\end{figure}	
\clearpage	
	
	\begin{figure}[h!]
		\includegraphics[width=0.32\textwidth,height=4cm]{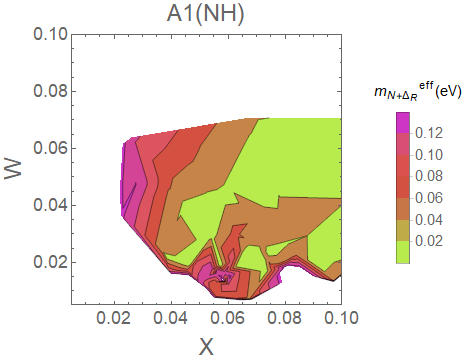}
		\includegraphics[width=0.32\textwidth,height=4cm]{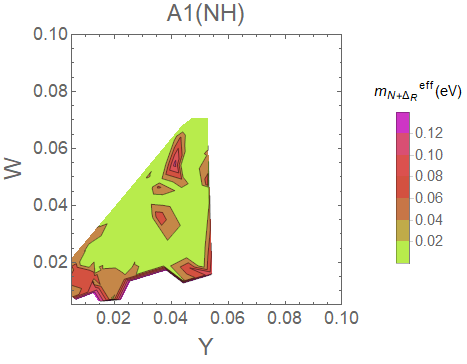}
	    \includegraphics[width=0.32\textwidth,height=4cm]{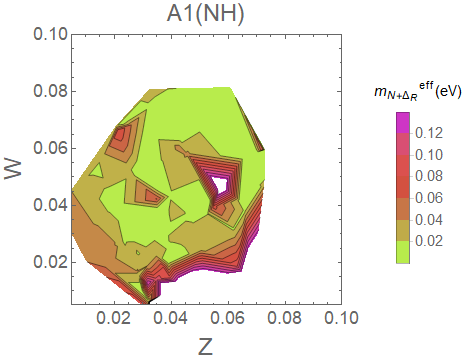}\\
        \includegraphics[width=0.32\textwidth,height=4cm]{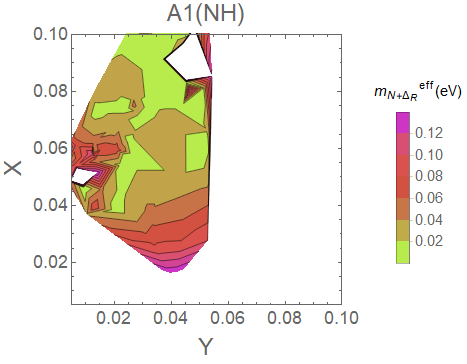}
        \includegraphics[width=0.32\textwidth,height=4cm]{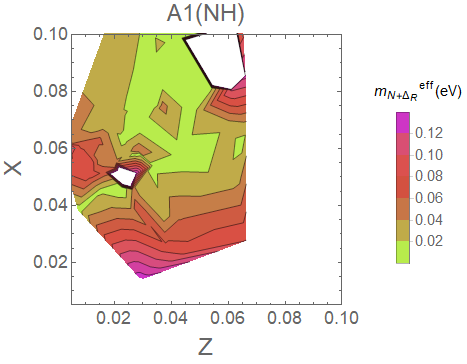}
        \includegraphics[width=0.32\textwidth,height=4cm]{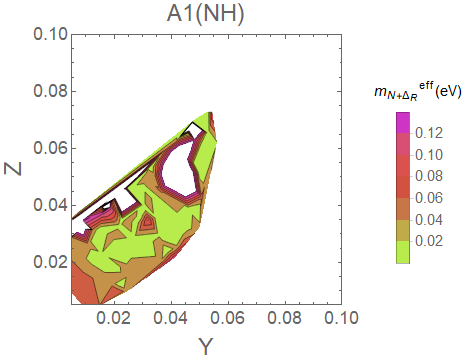}

		\caption{The various combinations of type II SS model parameters (in eV) with the new physics contribution to effective masseffective mass as the contour, where 0.061 eV is the KamLAND-Zen upper limit in A1 class of two texture zero neutrino mass matrix for normal hierarchy. } \label{fig6}
	\end{figure}
	
		\begin{figure}[h!]
		\includegraphics[width=0.32\textwidth,height=4cm]{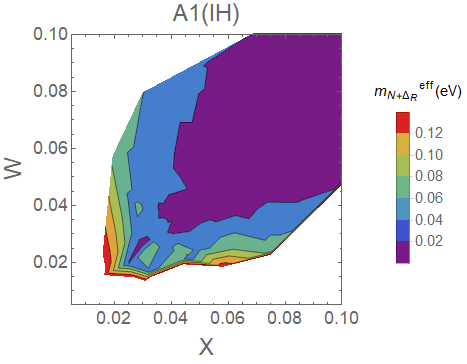}
		\includegraphics[width=0.32\textwidth,height=4cm]{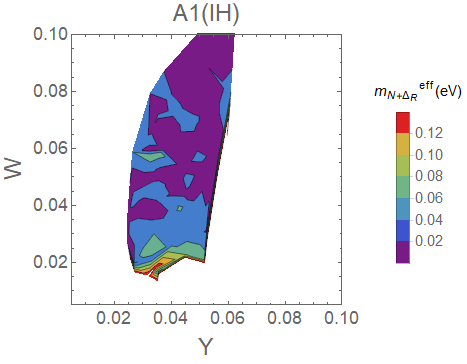}
		\includegraphics[width=0.32\textwidth,height=4cm]{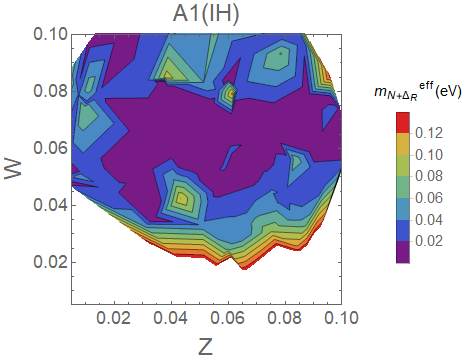}\\
		\includegraphics[width=0.32\textwidth,height=4cm]{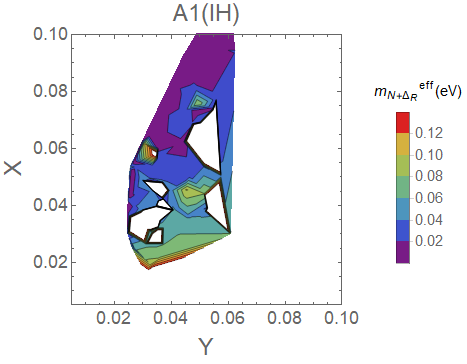}
		\includegraphics[width=0.32\textwidth,height=4cm]{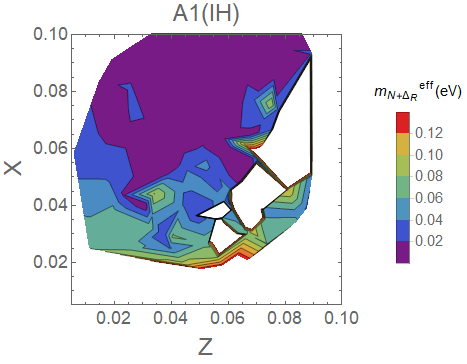}
		\includegraphics[width=0.32\textwidth,height=4cm]{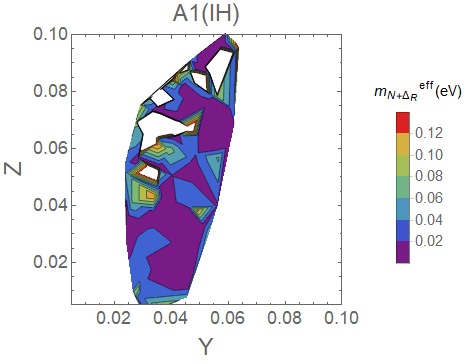}
		
		\caption{The various combinations of type II SS model parameters (in eV) with the new physics contribution to effective mass as the contour, where 0.061 eV is the KamLAND-Zen upper limit in A1 class of two texture zero neutrino mass matrix for inverted hierarchy.} \label{fig6}
	\end{figure}
	\clearpage		
	\begin{figure}[h!]
	\includegraphics[width=0.32\textwidth,height=4cm]{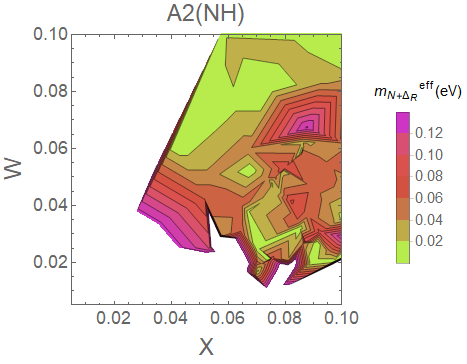}
	\includegraphics[width=0.32\textwidth,height=4cm]{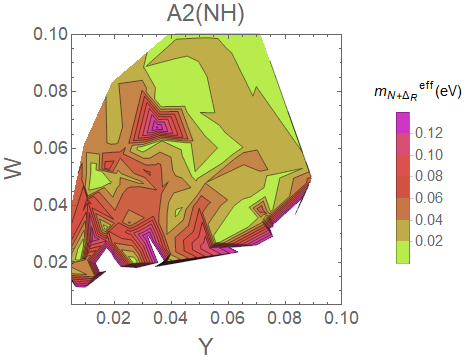}
	\includegraphics[width=0.32\textwidth,height=4cm]{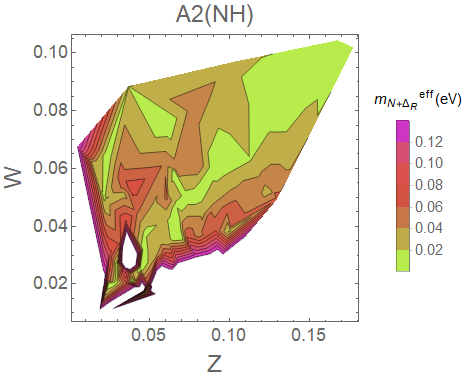}\\
	\includegraphics[width=0.32\textwidth,height=4cm]{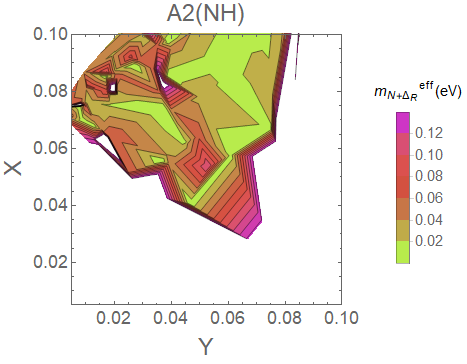}
	\includegraphics[width=0.32\textwidth,height=4cm]{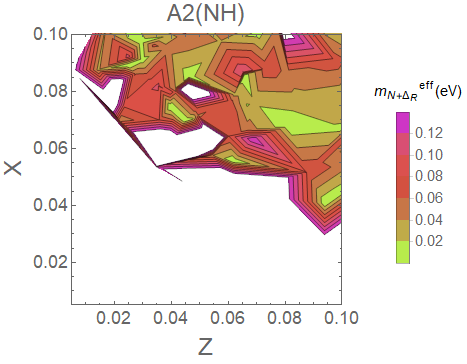}
	\includegraphics[width=0.32\textwidth,height=4cm]{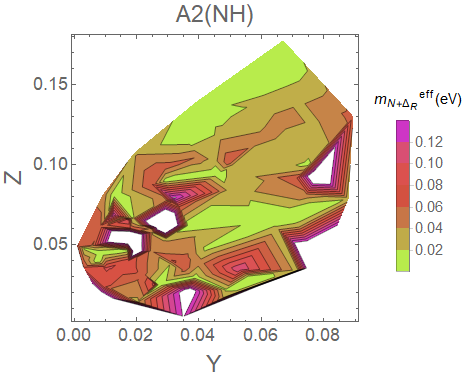}
	
	\caption{The various combinations of type II SS model parameters (in eV) with the new physics contribution to effective mass as the contour, where 0.061 eV is the KamLAND-Zen upper limit in A2 class of two texture zero neutrino mass matrix for normal hierarchy.} \label{fig6}
\end{figure}
	
\begin{figure}[h!]
	\includegraphics[width=0.32\textwidth,height=4cm]{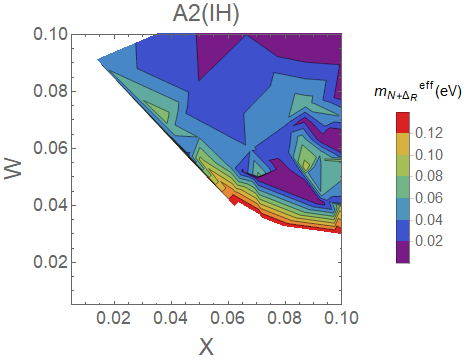}
	\includegraphics[width=0.32\textwidth,height=4cm]{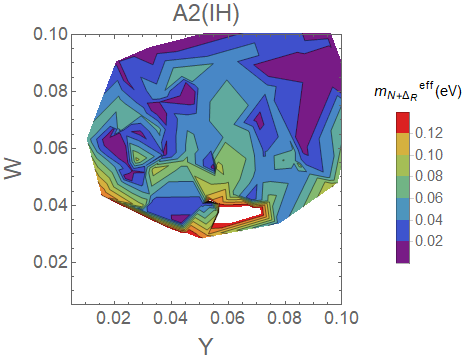}
	\includegraphics[width=0.32\textwidth,height=4cm]{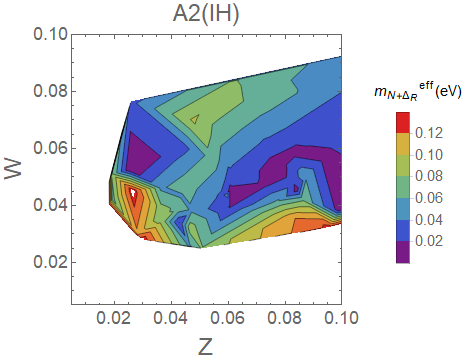}\\
	\includegraphics[width=0.32\textwidth,height=4cm]{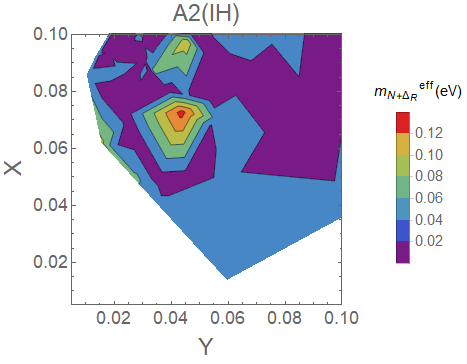}
	\includegraphics[width=0.32\textwidth,height=4cm]{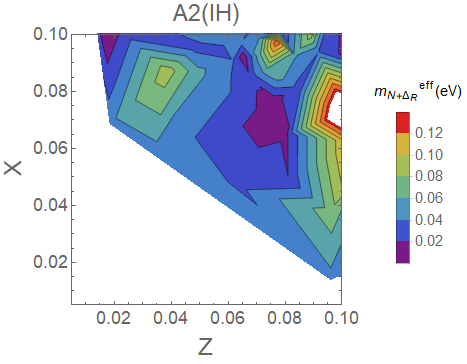}
	\includegraphics[width=0.32\textwidth,height=4cm]{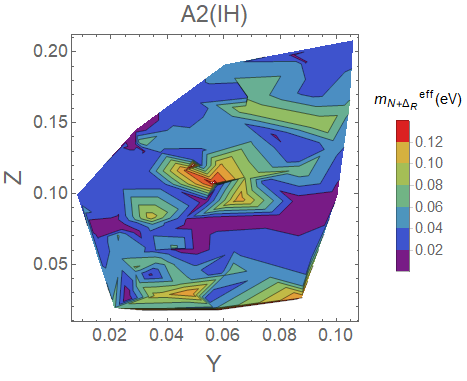}
	
	\caption{The various combinations of type II SS model parameters (in eV) with the new physics contribution to effective mass as the contour, where 0.061 eV is the KamLAND-Zen upper limit in A2 class of two texture zero neutrino mass matrix for inverted hierarchy.} \label{fig6}
\end{figure}	
	\clearpage	
	
\begin{figure}[h!]
	\includegraphics[width=0.32\textwidth,height=4cm]{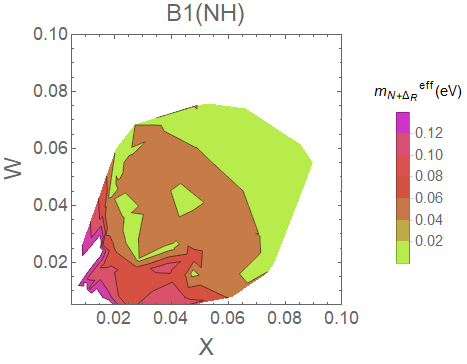}
	\includegraphics[width=0.32\textwidth,height=4cm]{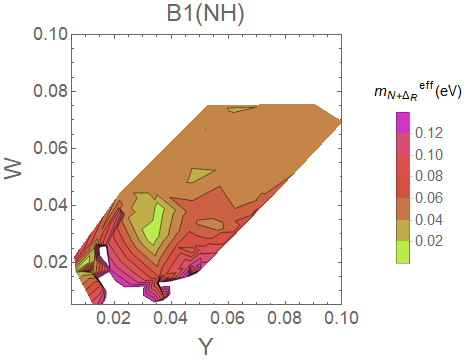}
	\includegraphics[width=0.32\textwidth,height=4cm]{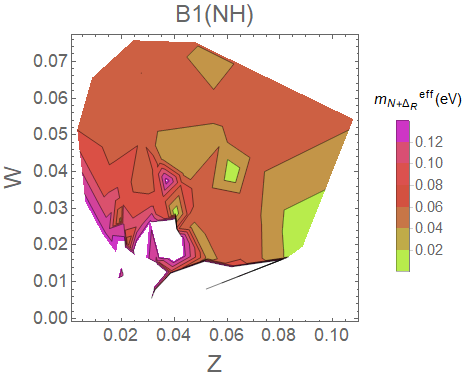}\\
	\includegraphics[width=0.32\textwidth,height=4cm]{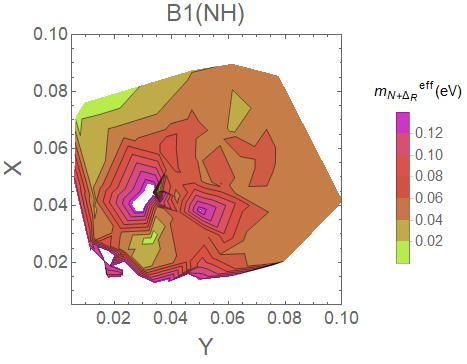}
	\includegraphics[width=0.32\textwidth,height=4cm]{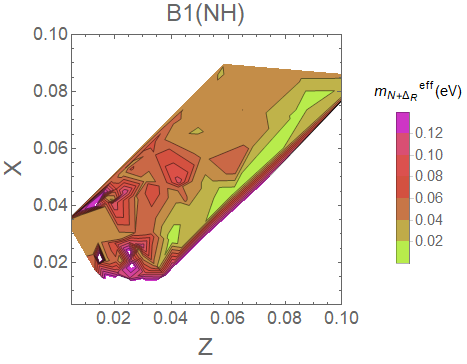}
	\includegraphics[width=0.32\textwidth,height=4cm]{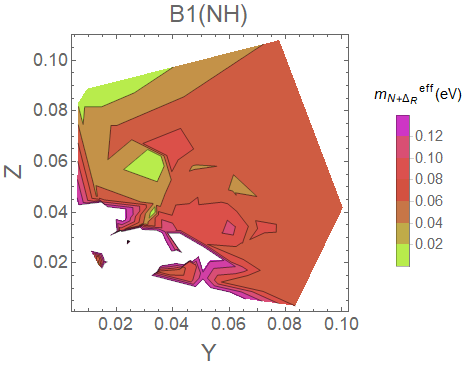}
	
	\caption{The various combinations of type II SS model parameters (in eV) with the new physics contribution to effective mass as the contour, where 0.061 eV is the KamLAND-Zen upper limit in B1 class of two texture zero neutrino mass matrix for normal hierarchy.} \label{fig6}
\end{figure}
	
\begin{figure}[h!]
	\includegraphics[width=0.32\textwidth,height=4cm]{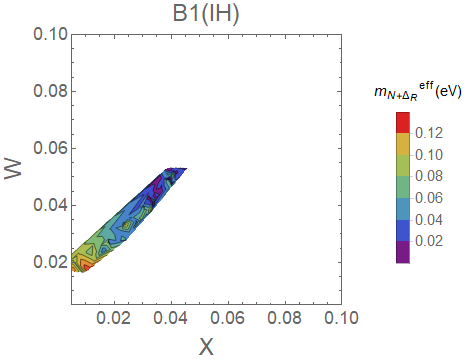}
	\includegraphics[width=0.32\textwidth,height=4cm]{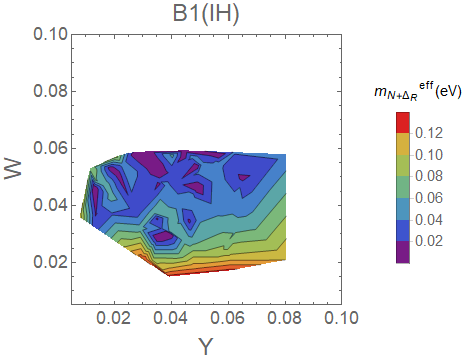}
	\includegraphics[width=0.32\textwidth,height=4cm]{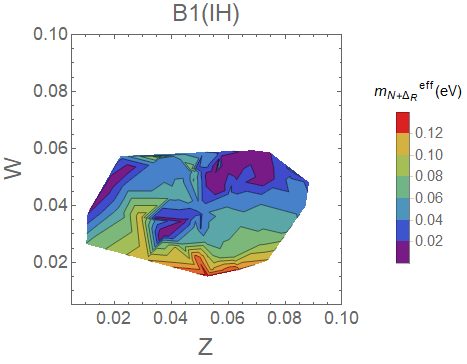}\\
	\includegraphics[width=0.32\textwidth,height=4cm]{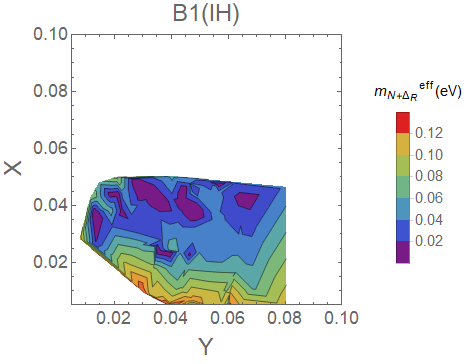}
	\includegraphics[width=0.32\textwidth,height=4cm]{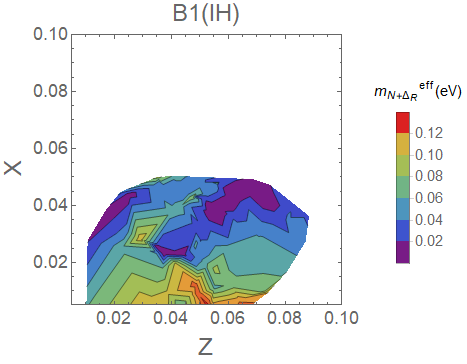}
	\includegraphics[width=0.32\textwidth,height=4cm]{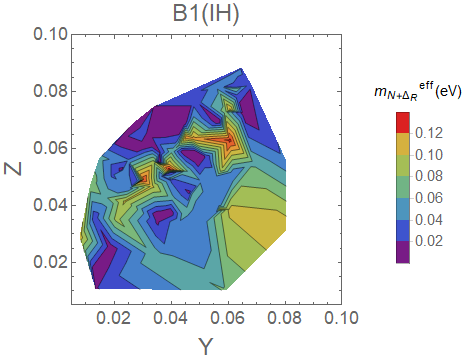}
	
	\caption{The various combinations of type II SS model parameters (in eV) with the new physics contribution to effective mass as the contour, where 0.061 eV is the KamLAND-Zen upper limit in B1 class of two texture zero neutrino mass matrix for inverted hierarchy.} \label{fig6}
\end{figure}				
	\clearpage
\begin{figure}[h!]
	\includegraphics[width=0.32\textwidth,height=4cm]{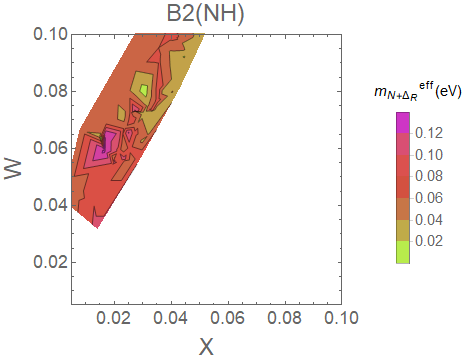}
	\includegraphics[width=0.32\textwidth,height=4cm]{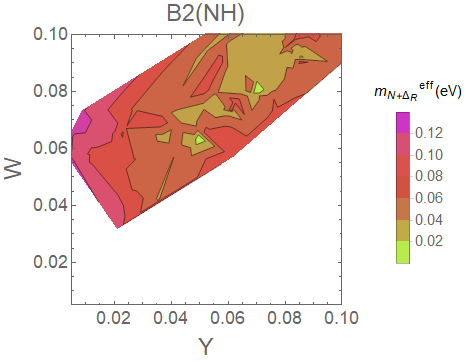}
	\includegraphics[width=0.32\textwidth,height=4cm]{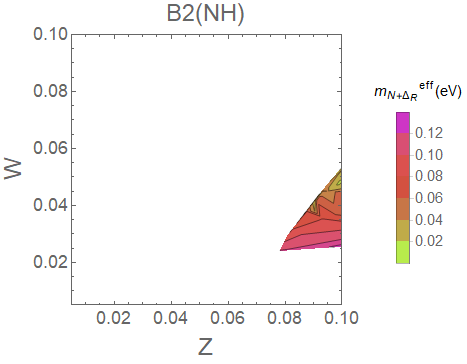}\\
	\includegraphics[width=0.32\textwidth,height=4cm]{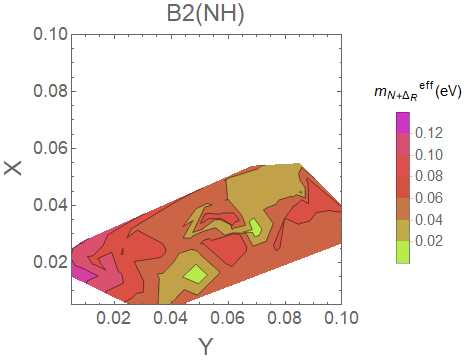}
	\includegraphics[width=0.32\textwidth,height=4cm]{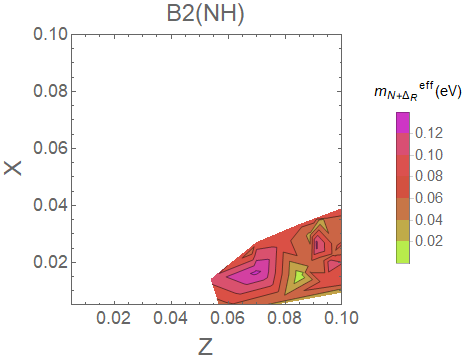}
	\includegraphics[width=0.32\textwidth,height=4cm]{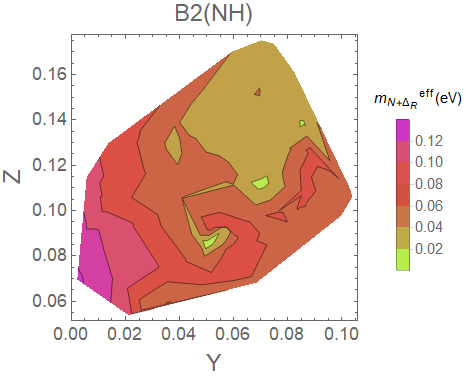}
	
	\caption{The various combinations of type II SS model parameters (in eV) with the new physics contribution to effective mass as the contour, where 0.061 eV is the KamLAND-Zen upper limit in B2 class of two texture zero neutrino mass matrix for normal hierarchy.} \label{fig6}
\end{figure}
		
\begin{figure}[h!]
	\includegraphics[width=0.32\textwidth,height=4cm]{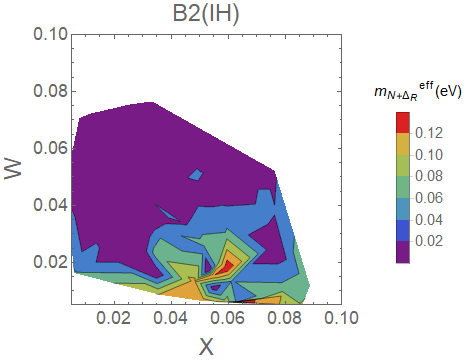}
	\includegraphics[width=0.32\textwidth,height=4cm]{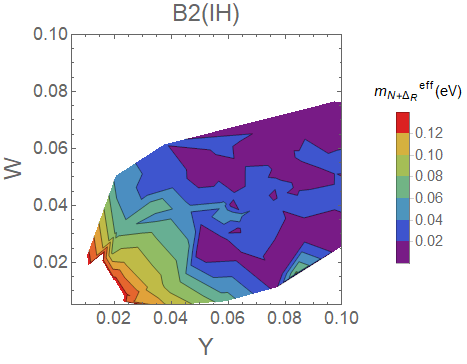}
	\includegraphics[width=0.32\textwidth,height=4cm]{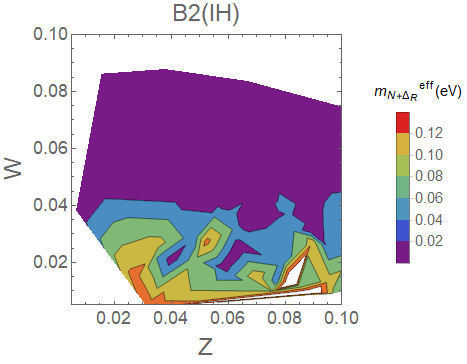}\\
	\includegraphics[width=0.32\textwidth,height=4cm]{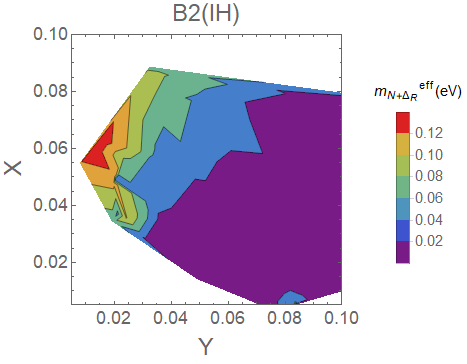}
	\includegraphics[width=0.32\textwidth,height=4cm]{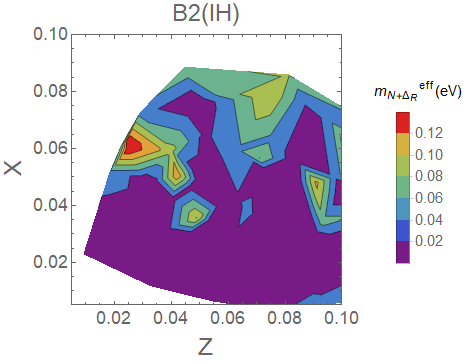}
	\includegraphics[width=0.32\textwidth,height=4cm]{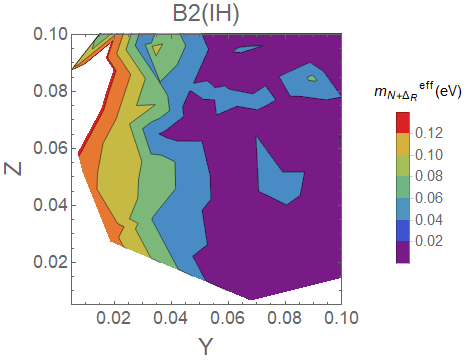}
	
	\caption{The various combinations of type II SS model parameters (in eV) with the new physics contribution to effective mass as the contour, where 0.061 eV is the KamLAND-Zen upper limit in B2 class of two texture zero neutrino mass matrix for inverted hierarchy.} \label{fig6}
\end{figure}	
\clearpage					
\begin{figure}[h!]
	\includegraphics[width=0.32\textwidth,height=4cm]{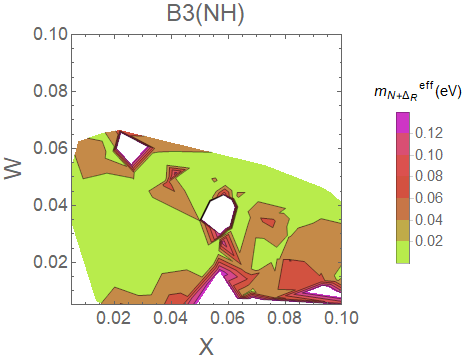}
	\includegraphics[width=0.32\textwidth,height=4cm]{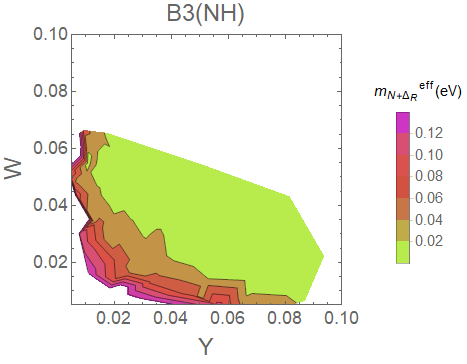}
	\includegraphics[width=0.32\textwidth,height=4cm]{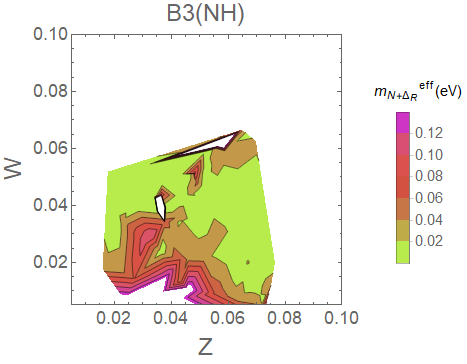}\\
	\includegraphics[width=0.32\textwidth,height=4cm]{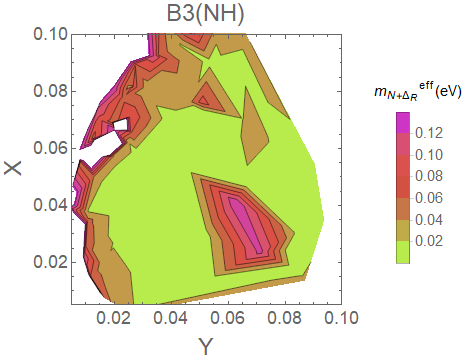}
	\includegraphics[width=0.32\textwidth,height=4cm]{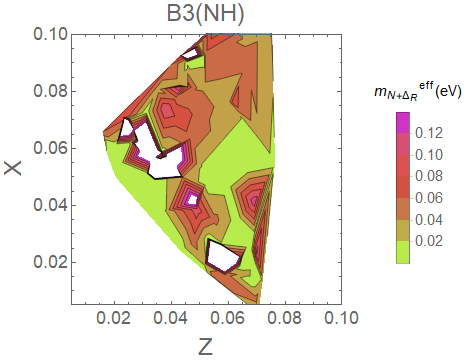}
	\includegraphics[width=0.32\textwidth,height=4cm]{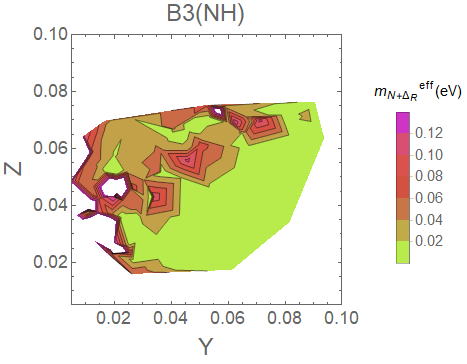}
	
	\caption{The various combinations of type II SS model parameters (in eV) with the new physics contribution to effective mass as the contour, where 0.061 eV is the KamLAND-Zen upper limit in B3 class of two texture zero neutrino mass matrix for normal hierarchy.} \label{fig6}
\end{figure}
		
\begin{figure}[h!]
	\includegraphics[width=0.32\textwidth,height=4cm]{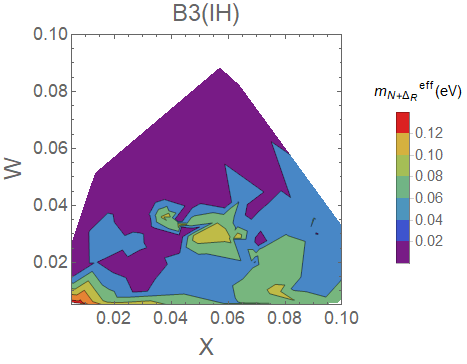}
	\includegraphics[width=0.32\textwidth,height=4cm]{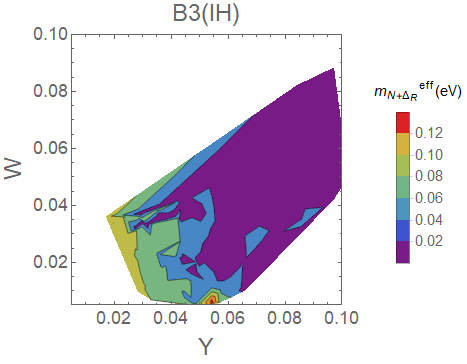}
	\includegraphics[width=0.32\textwidth,height=4cm]{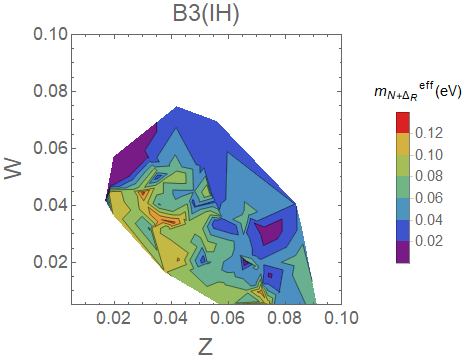}\\
	\includegraphics[width=0.32\textwidth,height=4cm]{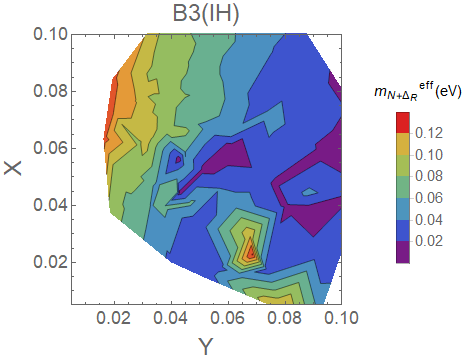}
	\includegraphics[width=0.32\textwidth,height=4cm]{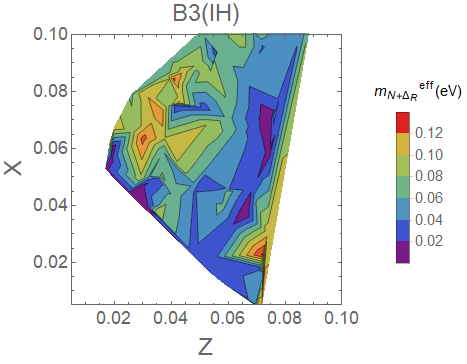}
	\includegraphics[width=0.32\textwidth,height=4cm]{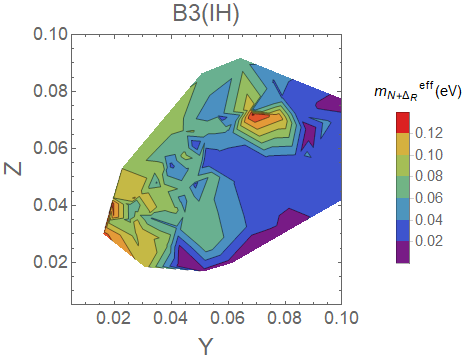}
	
	\caption{The various combinations of type II SS model parameters (in eV) with the new physics contribution to effective mass as the contour, where 0.061 eV is the KamLAND-Zen upper limit in B3 class of two texture zero neutrino mass matrix for inverted hierarchy.} \label{fig6}
\end{figure}
\clearpage			
\begin{figure}[h!]
	\includegraphics[width=0.32\textwidth,height=4cm]{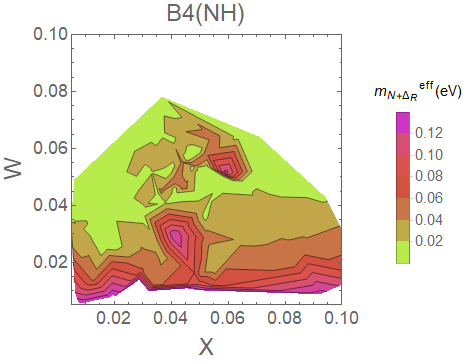}
	\includegraphics[width=0.32\textwidth,height=4cm]{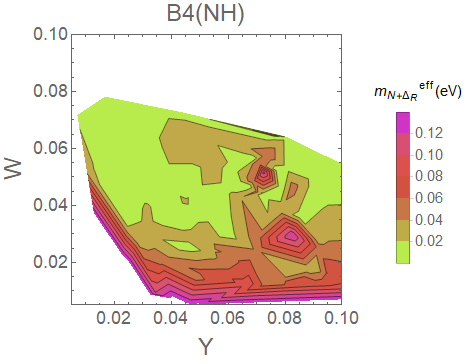}
	\includegraphics[width=0.32\textwidth,height=4cm]{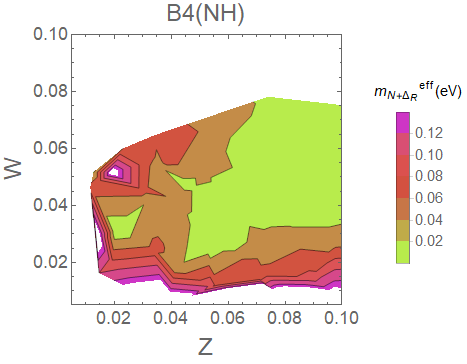}\\
	\includegraphics[width=0.32\textwidth,height=4cm]{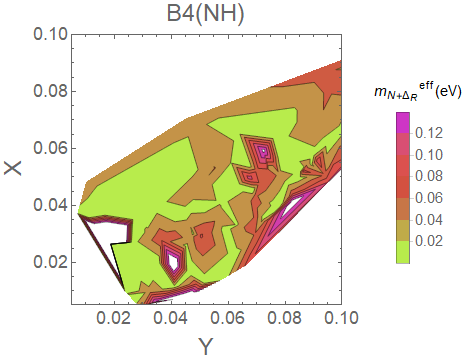}
	\includegraphics[width=0.32\textwidth,height=4cm]{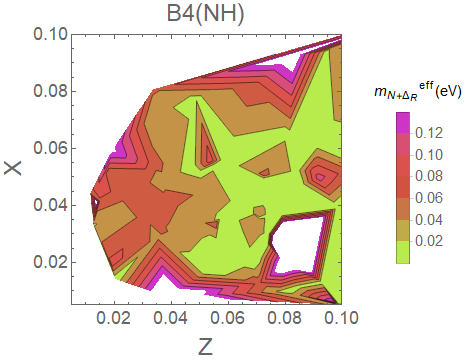}
	\includegraphics[width=0.32\textwidth,height=4cm]{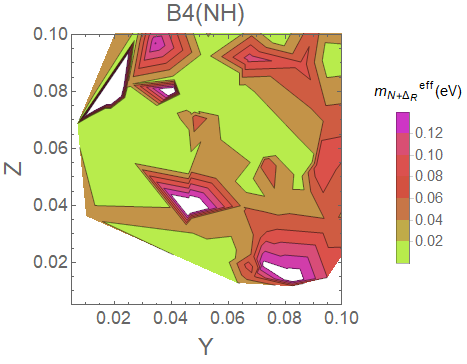}
	
	\caption{The various combinations of type II SS model parameters (in eV) with the new physics contribution to effective mass as the contour, where 0.061 eV is the KamLAND-Zen upper limit in B4 class of two texture zero neutrino mass matrix for normal hierarchy.} \label{fig6}
\end{figure}
\begin{figure}[h!]
	\includegraphics[width=0.32\textwidth,height=4cm]{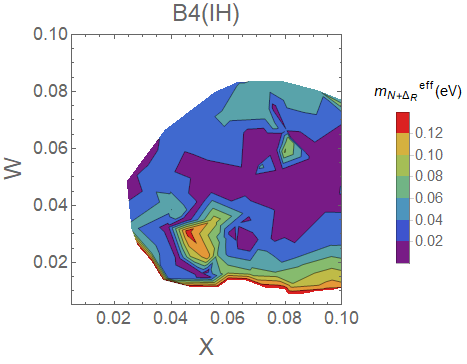}
	\includegraphics[width=0.32\textwidth,height=4cm]{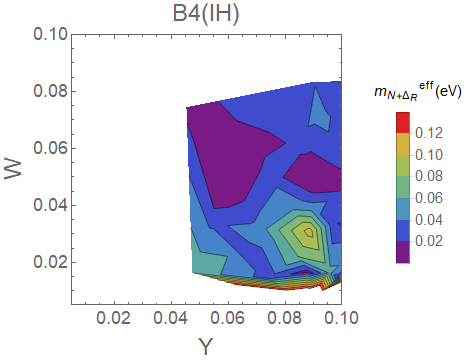}
	\includegraphics[width=0.32\textwidth,height=4cm]{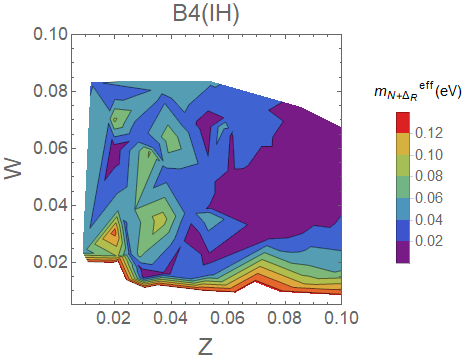}\\
	\includegraphics[width=0.32\textwidth,height=4cm]{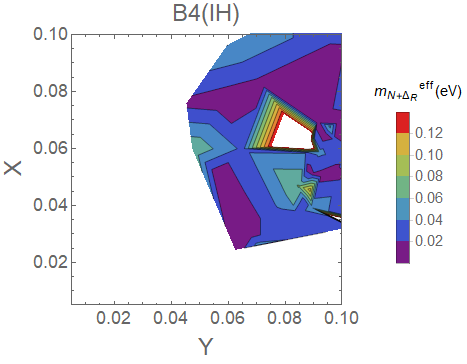}
	\includegraphics[width=0.32\textwidth,height=4cm]{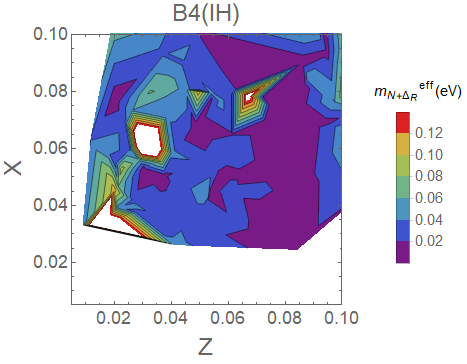}
	\includegraphics[width=0.32\textwidth,height=4cm]{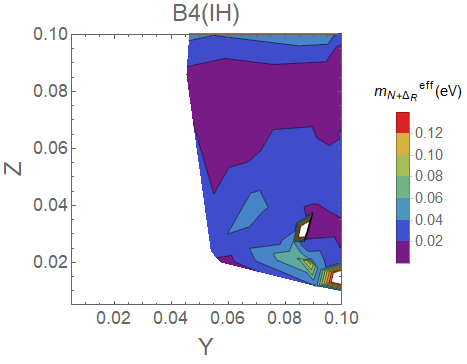}
	
	\caption{The various combinations of type II SS model parameters (in eV) with the new physics contribution to effective mass as the contour, where 0.061 eV is the KamLAND-Zen upper limit in B4 class of two texture zero neutrino mass matrix for inverted hierarchy.} \label{fig6}
\end{figure}
	
\begin{figure}[h!]
		\includegraphics[width=0.32\textwidth,height=4cm]{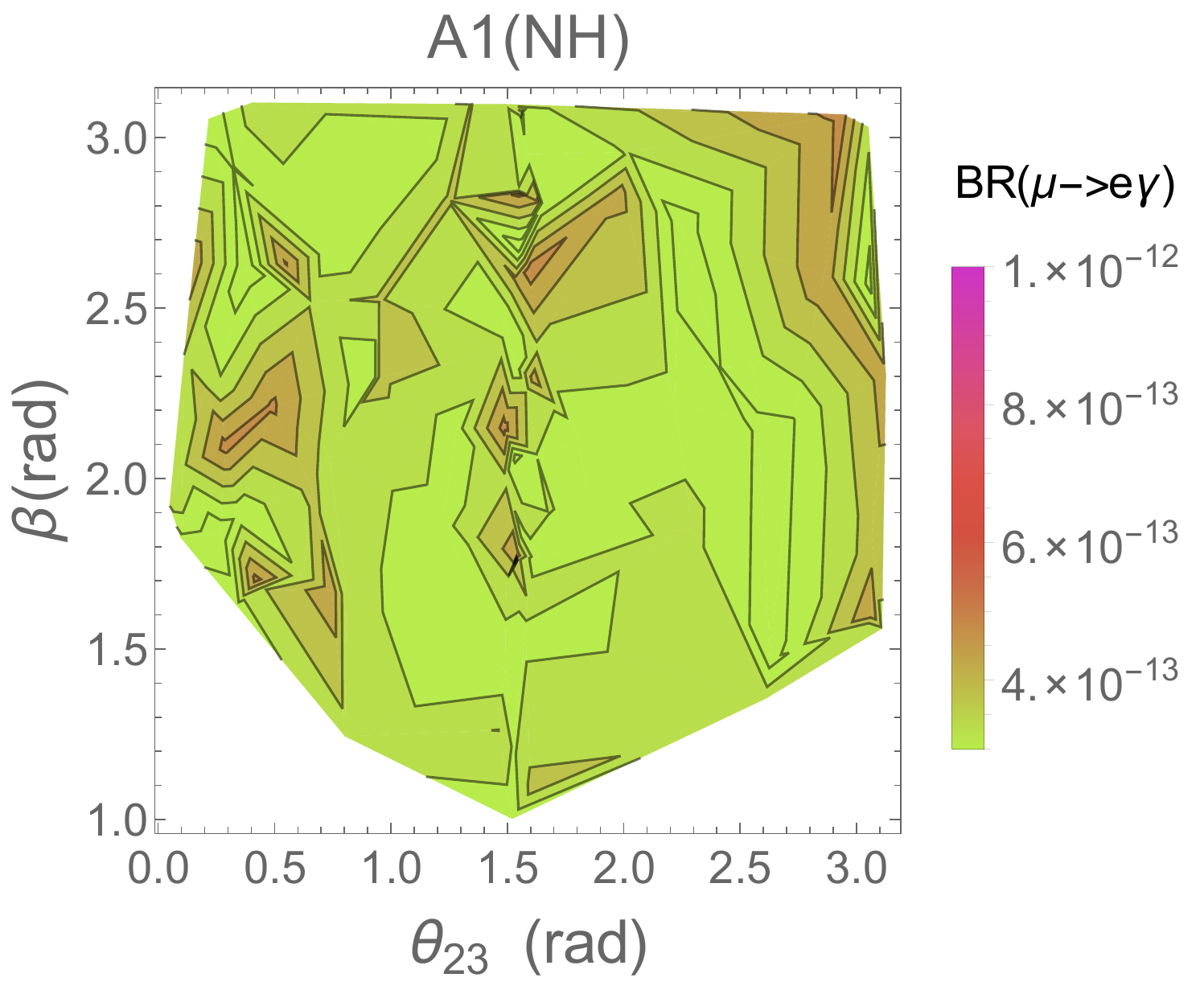}
	\includegraphics[width=0.32\textwidth,height=4cm]{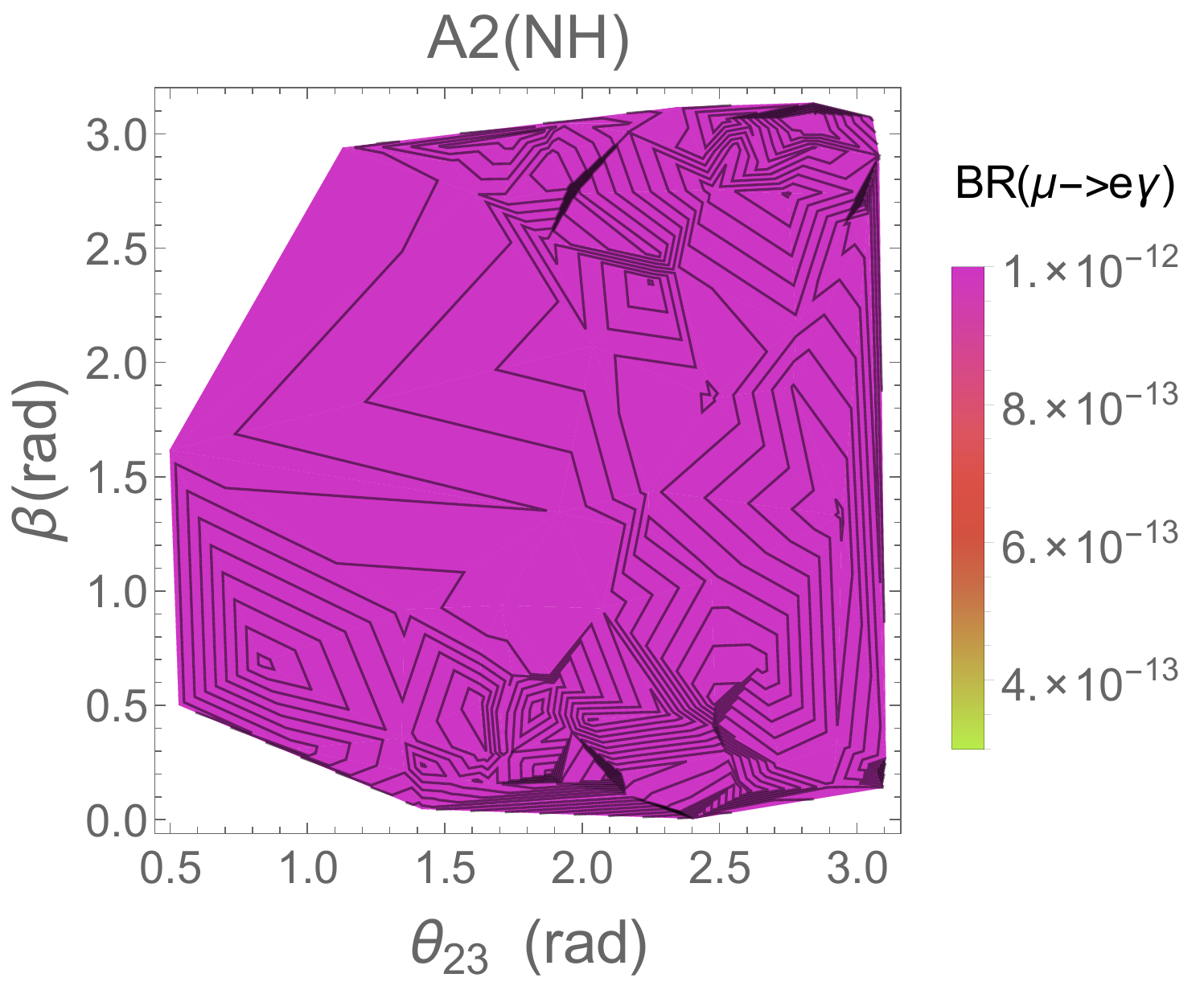}
	\includegraphics[width=0.32\textwidth,height=4cm]{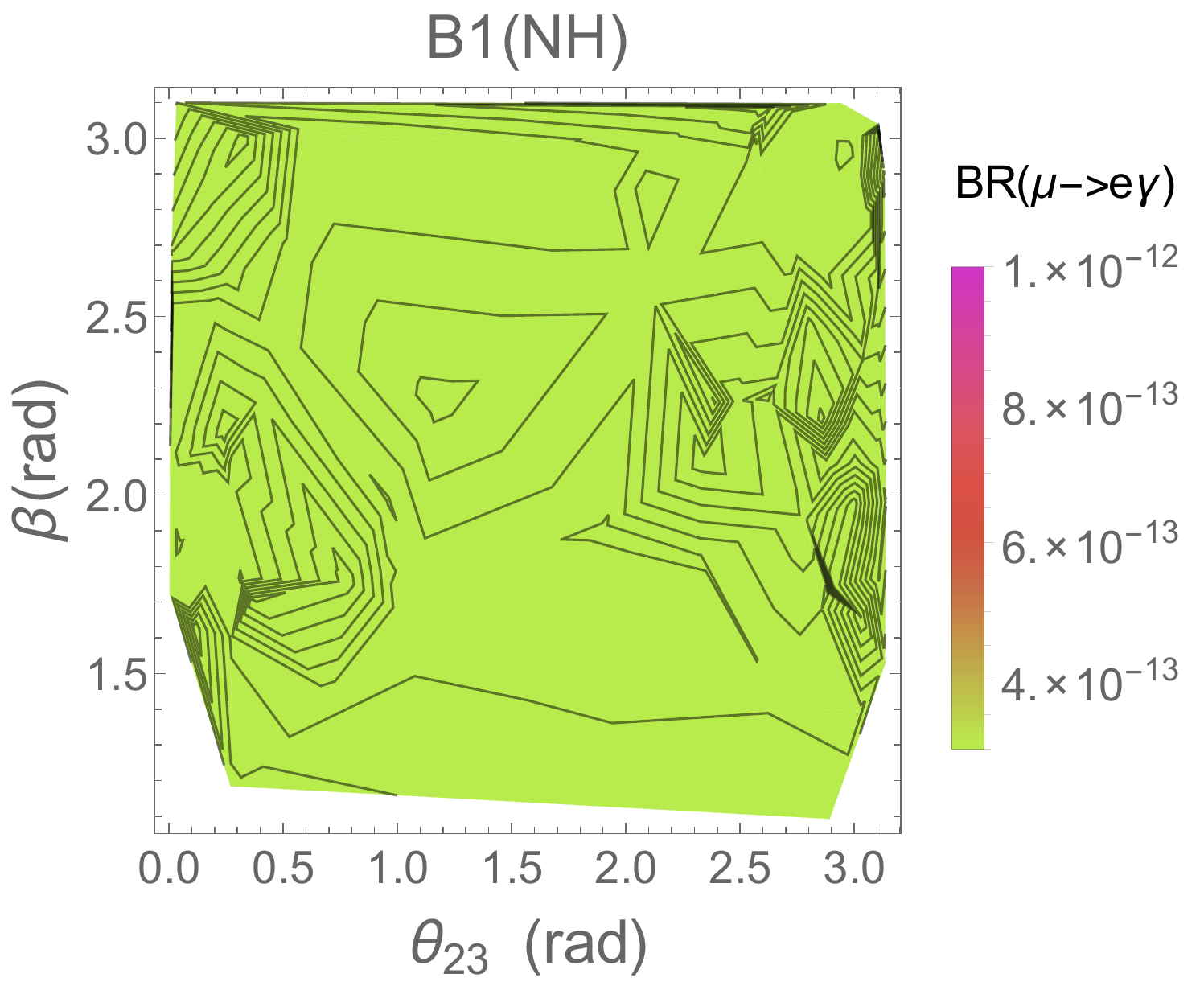}\\
	\includegraphics[width=0.32\textwidth,height=4cm]{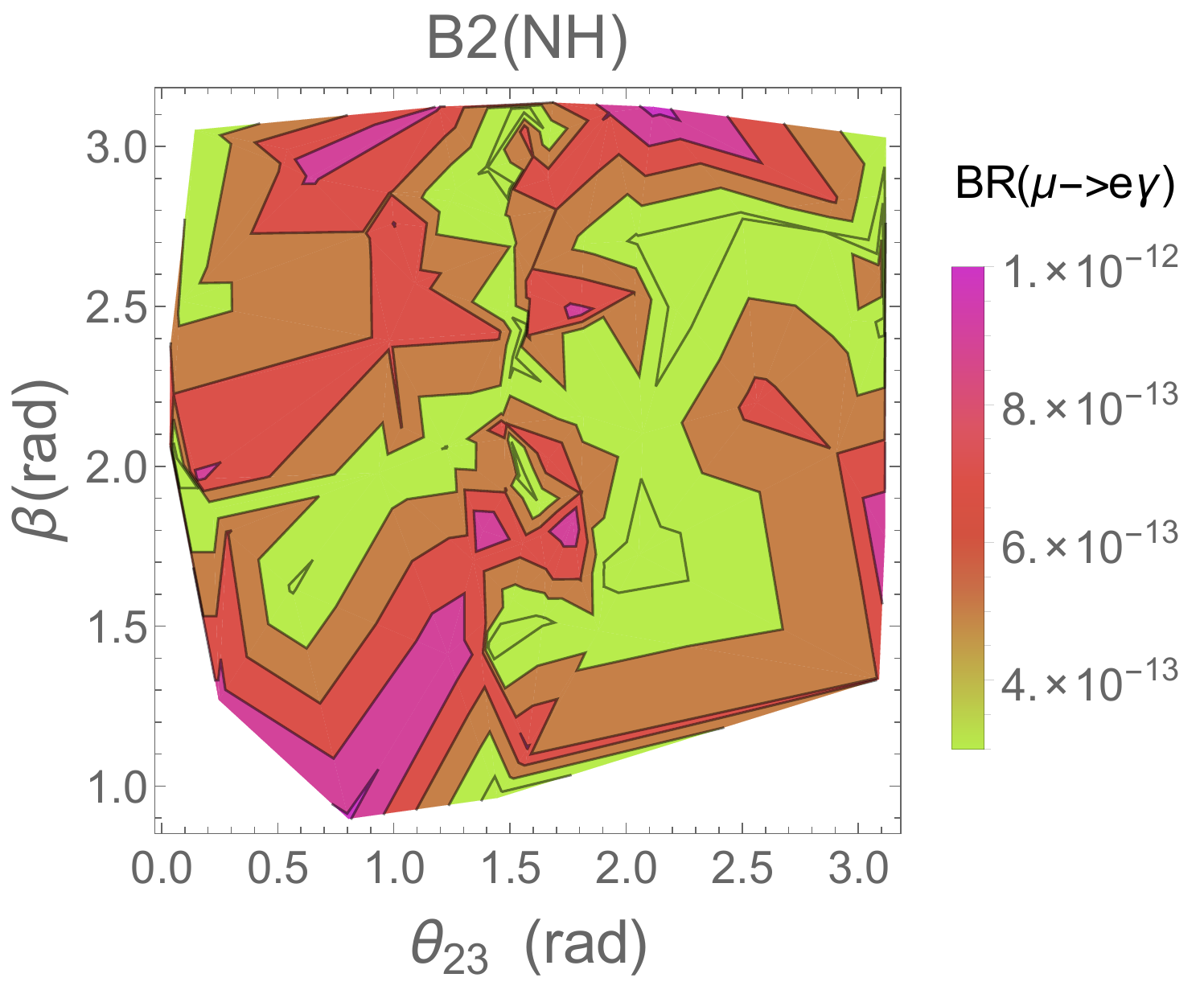}
	\includegraphics[width=0.32\textwidth,height=4cm]{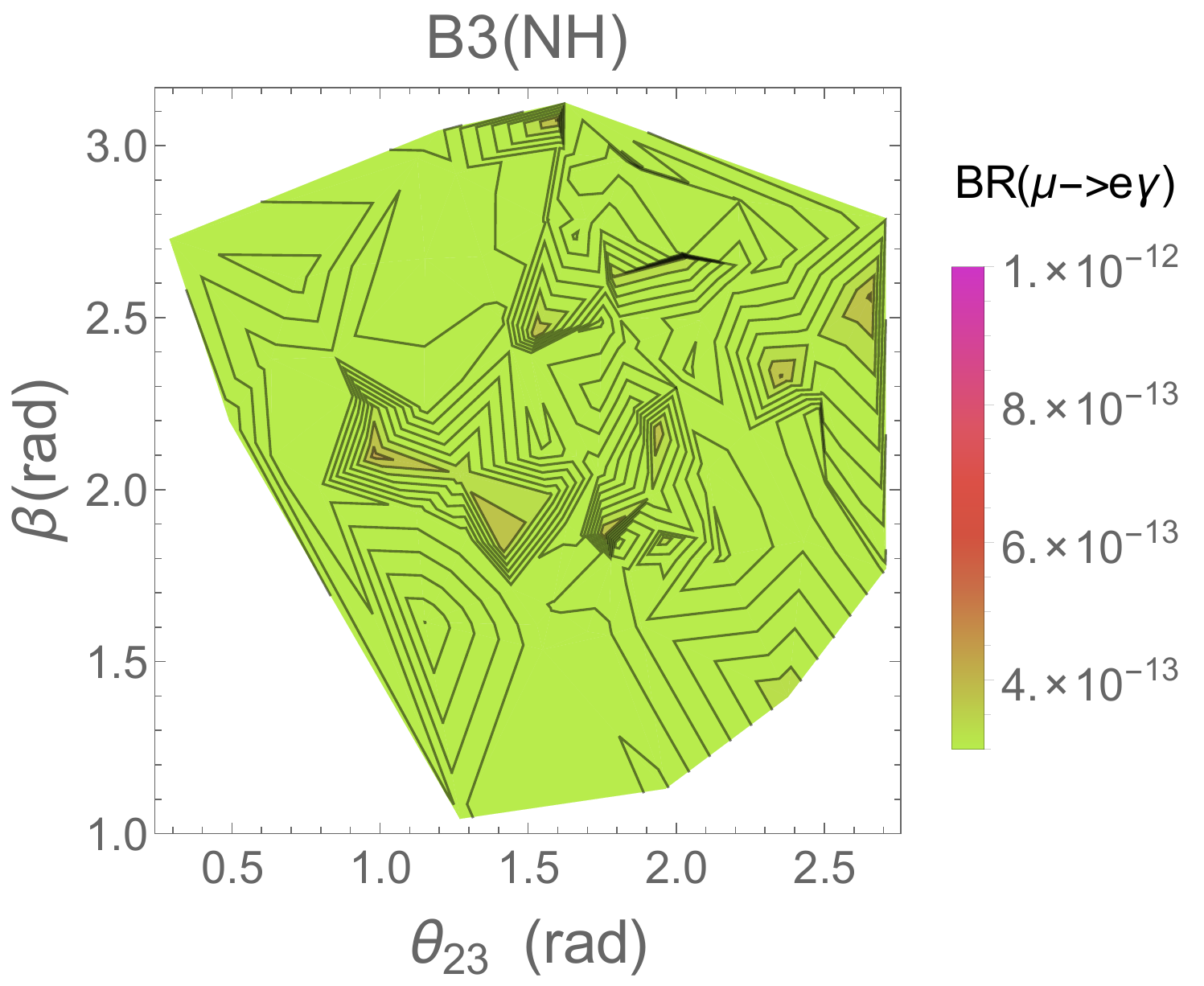}
	\includegraphics[width=0.32\textwidth,height=4cm]{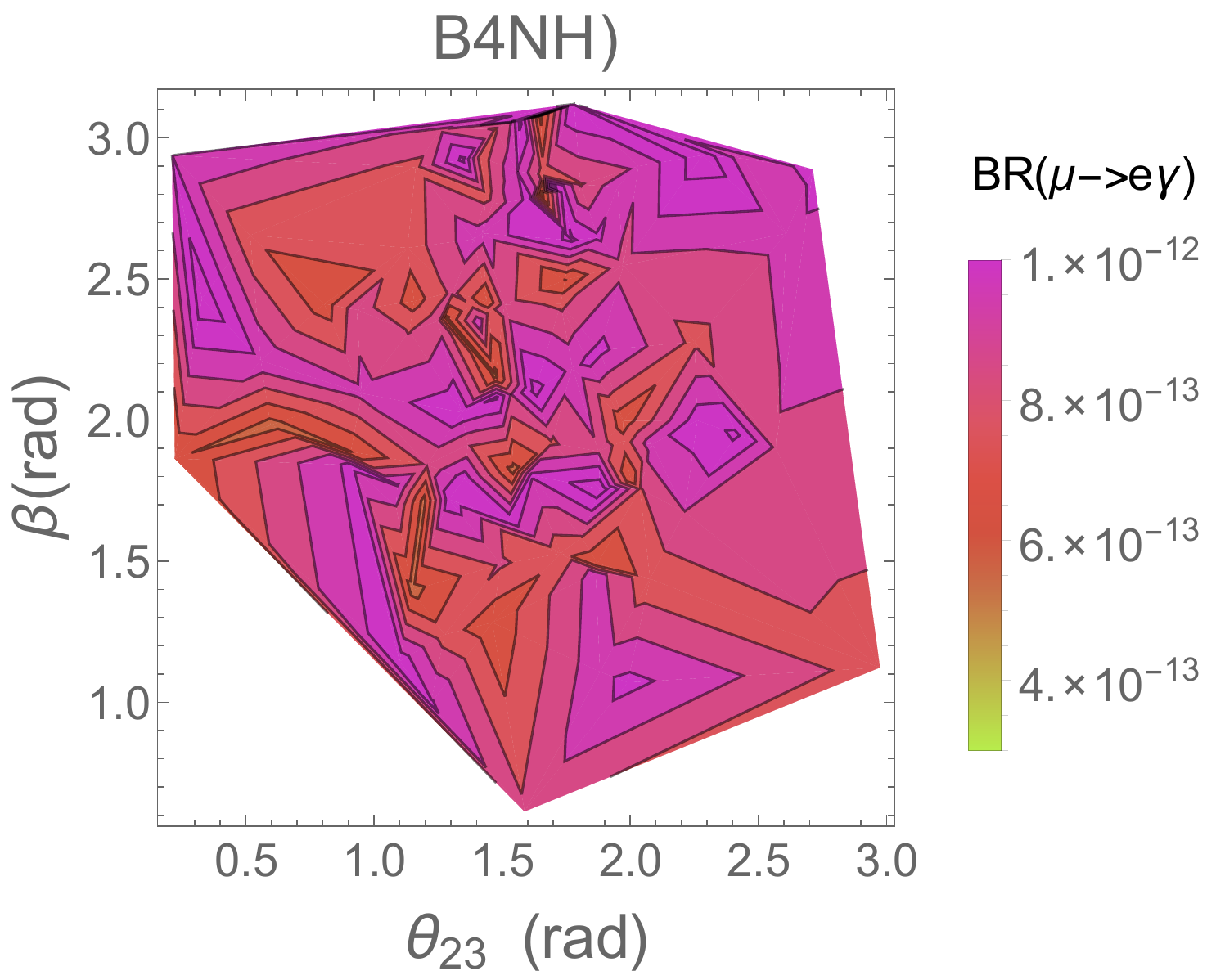}
	
	\caption{Atmospheric mixing angle, $\theta_{23}$ Vs Majorana phase $\alpha$ (for NH) with BR for $\rm \mu\rightarrow e\gamma $ as the contour where $4.2\times 10^{-13}$ is the upperlimit for BR given by MEG experiment. } \label{fig6}
\end{figure}				
	
\begin{figure}[h!]
	\includegraphics[width=0.32\textwidth,height=4cm]{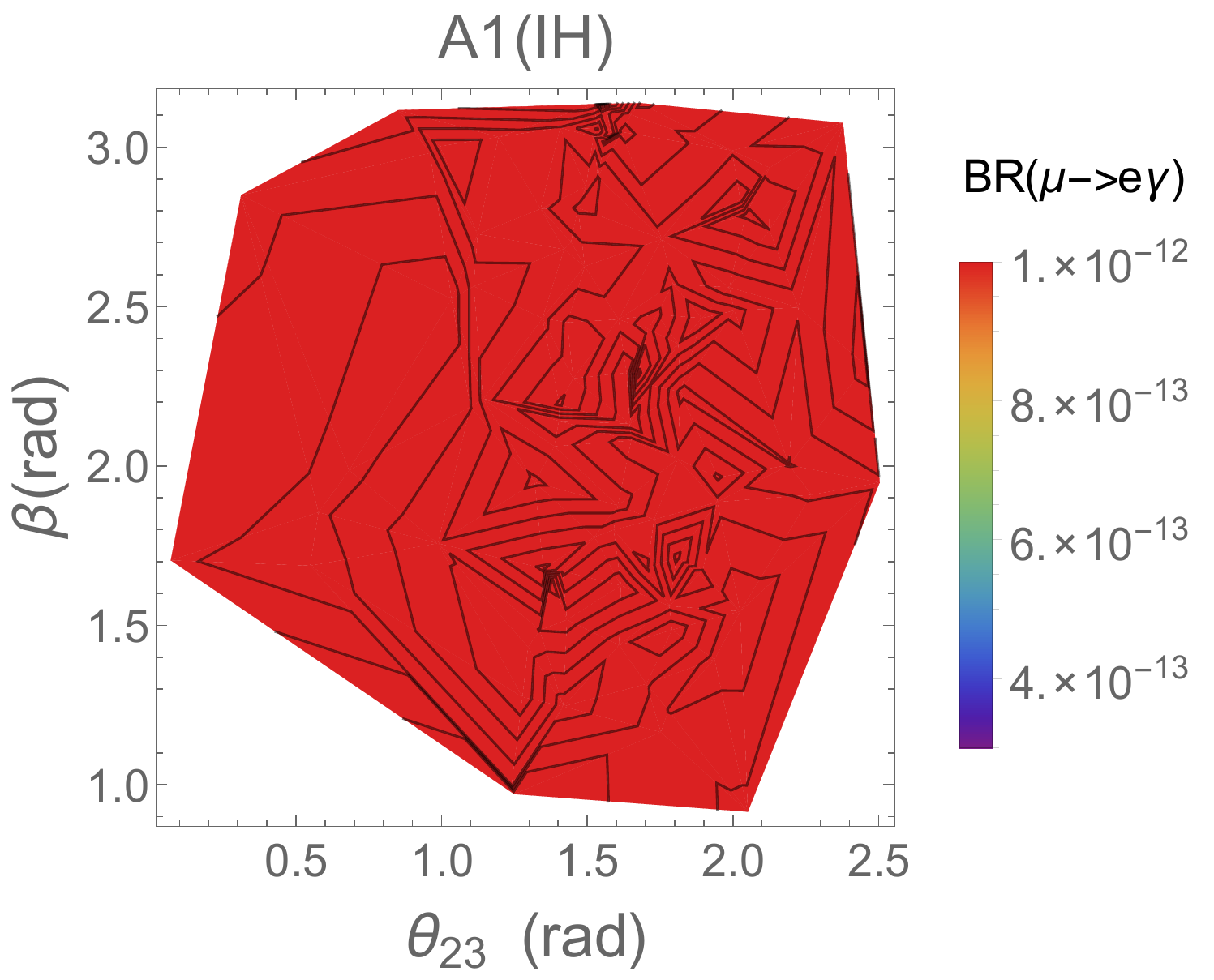}
	\includegraphics[width=0.32\textwidth,height=4cm]{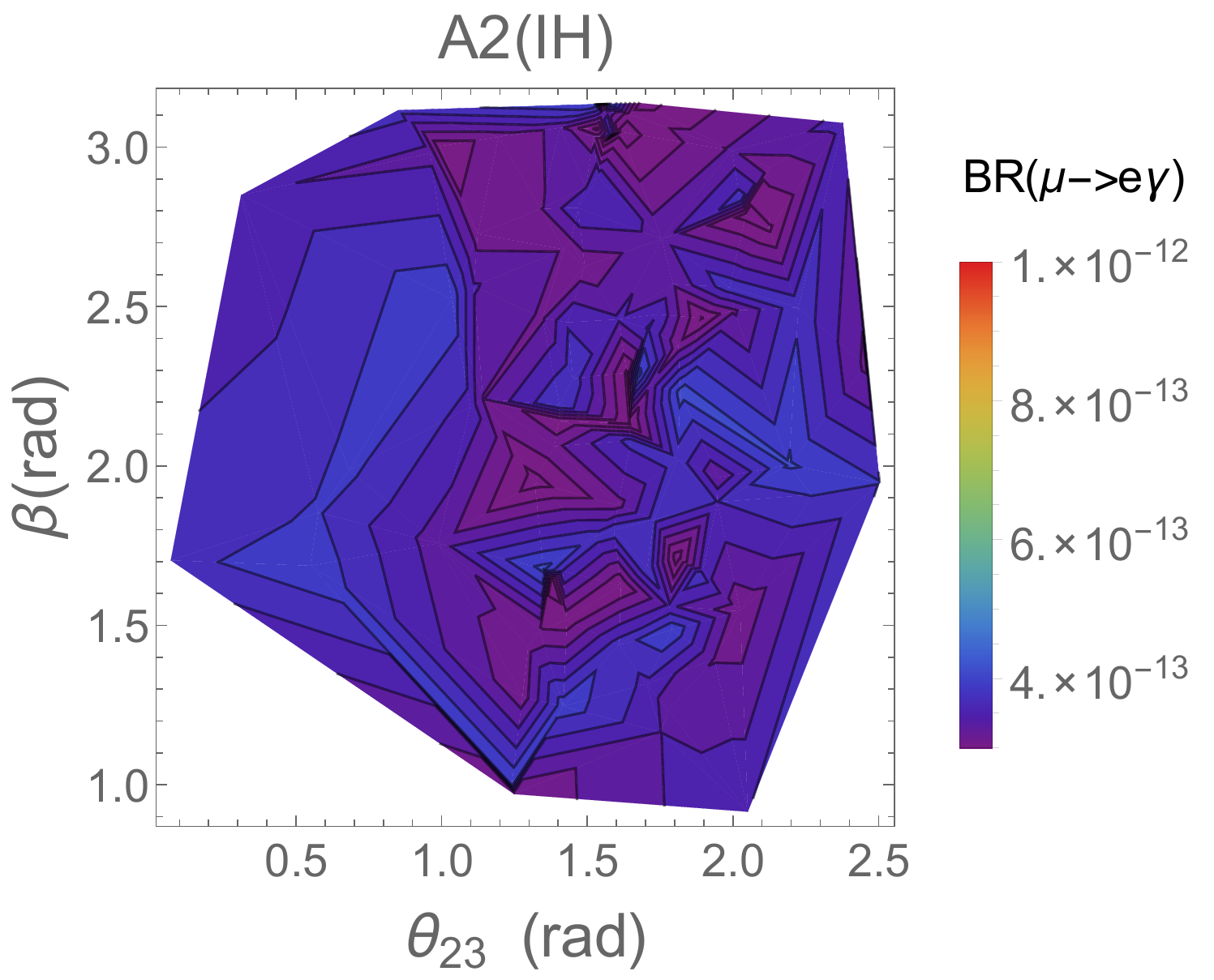}
	\includegraphics[width=0.32\textwidth,height=4cm]{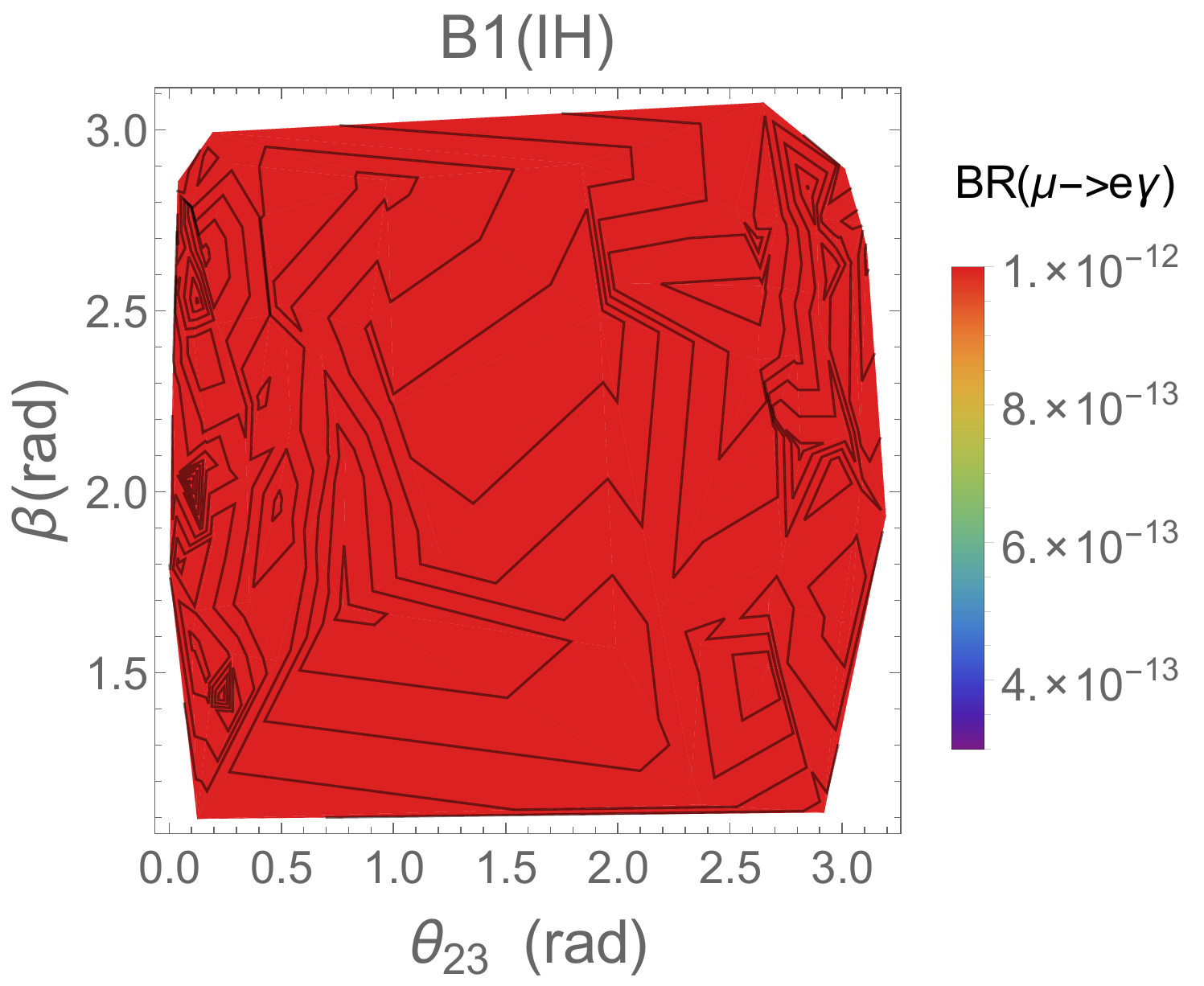}\\
	\includegraphics[width=0.32\textwidth,height=4cm]{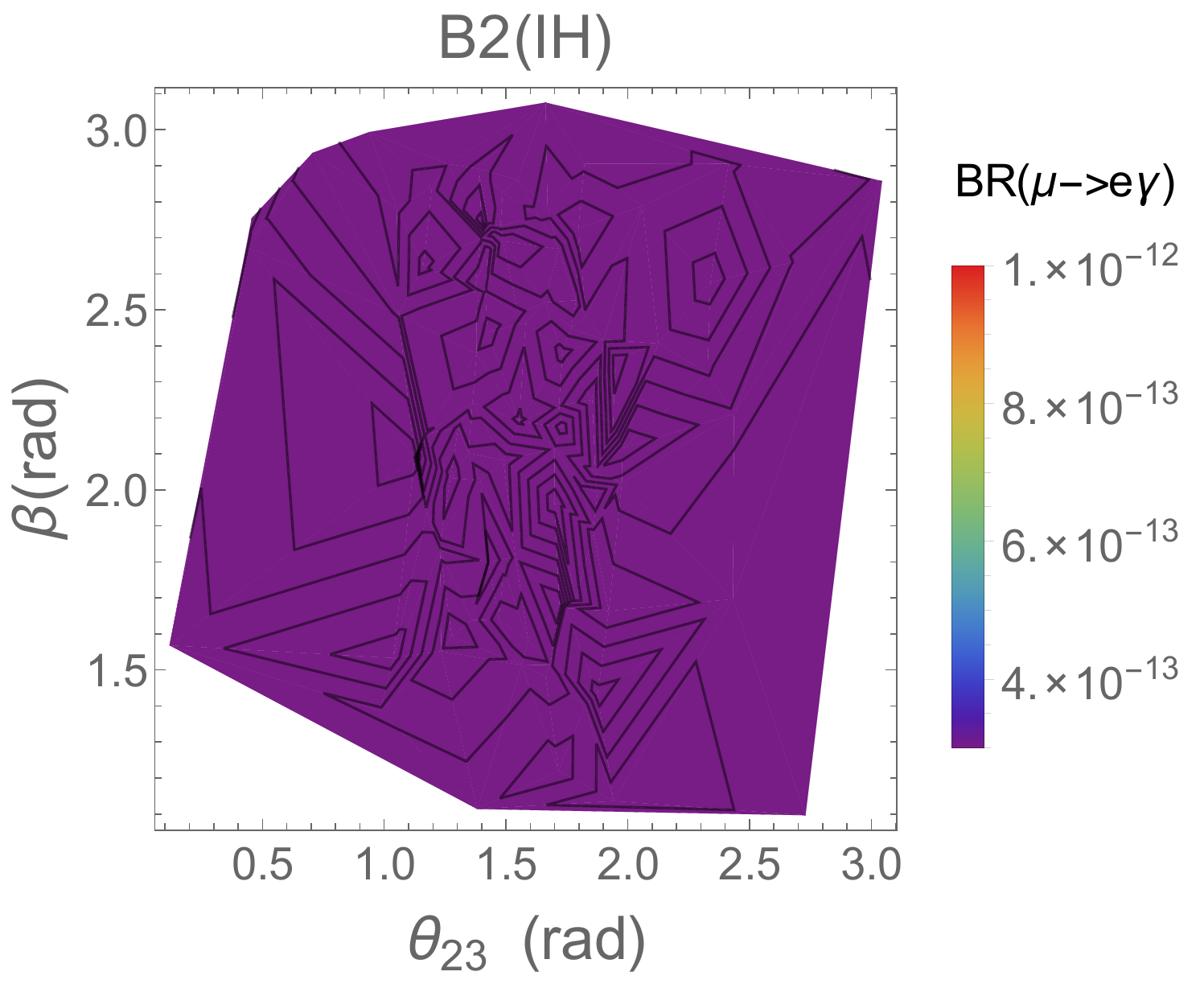}
	\includegraphics[width=0.32\textwidth,height=4cm]{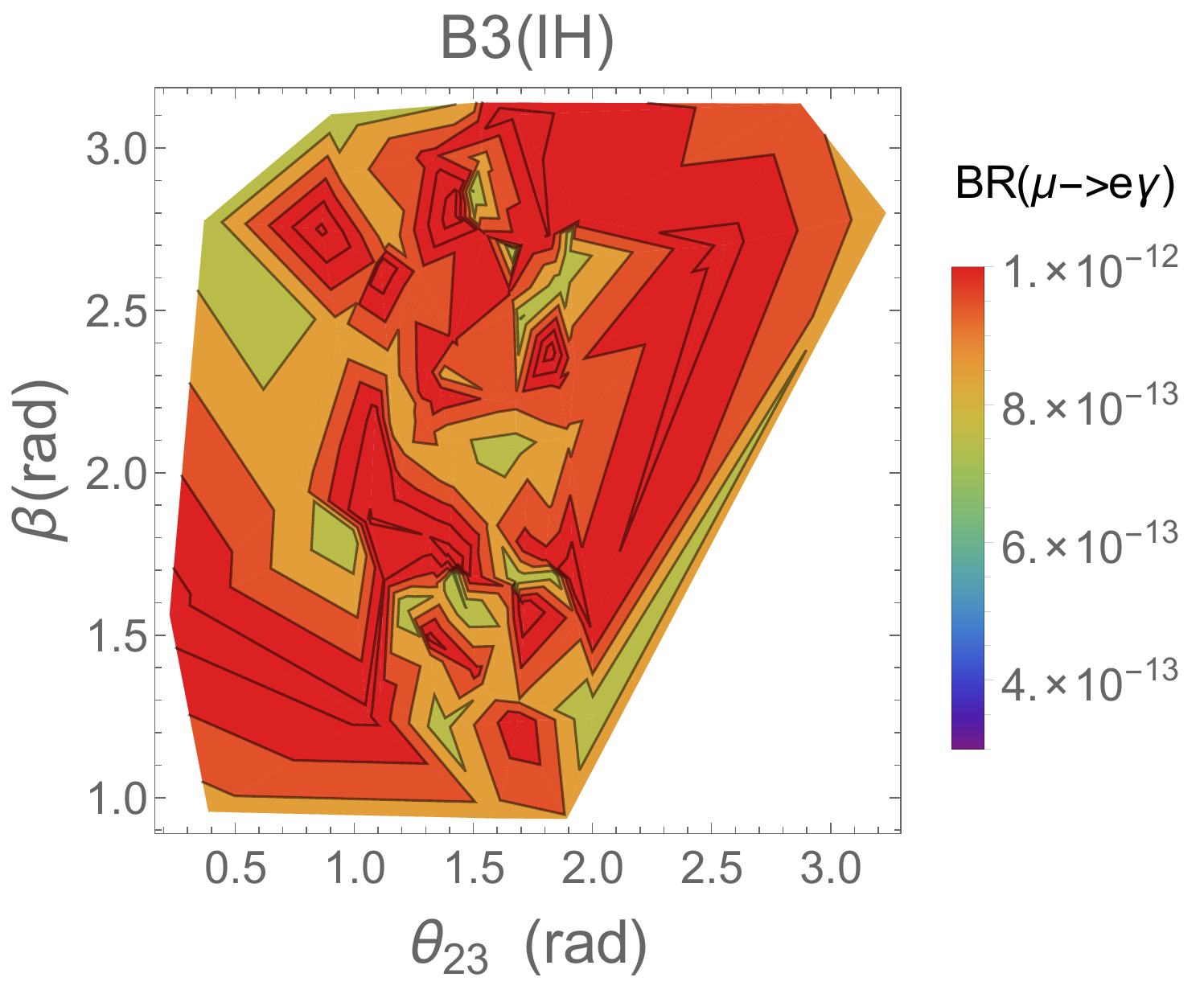}
	\includegraphics[width=0.32\textwidth,height=4cm]{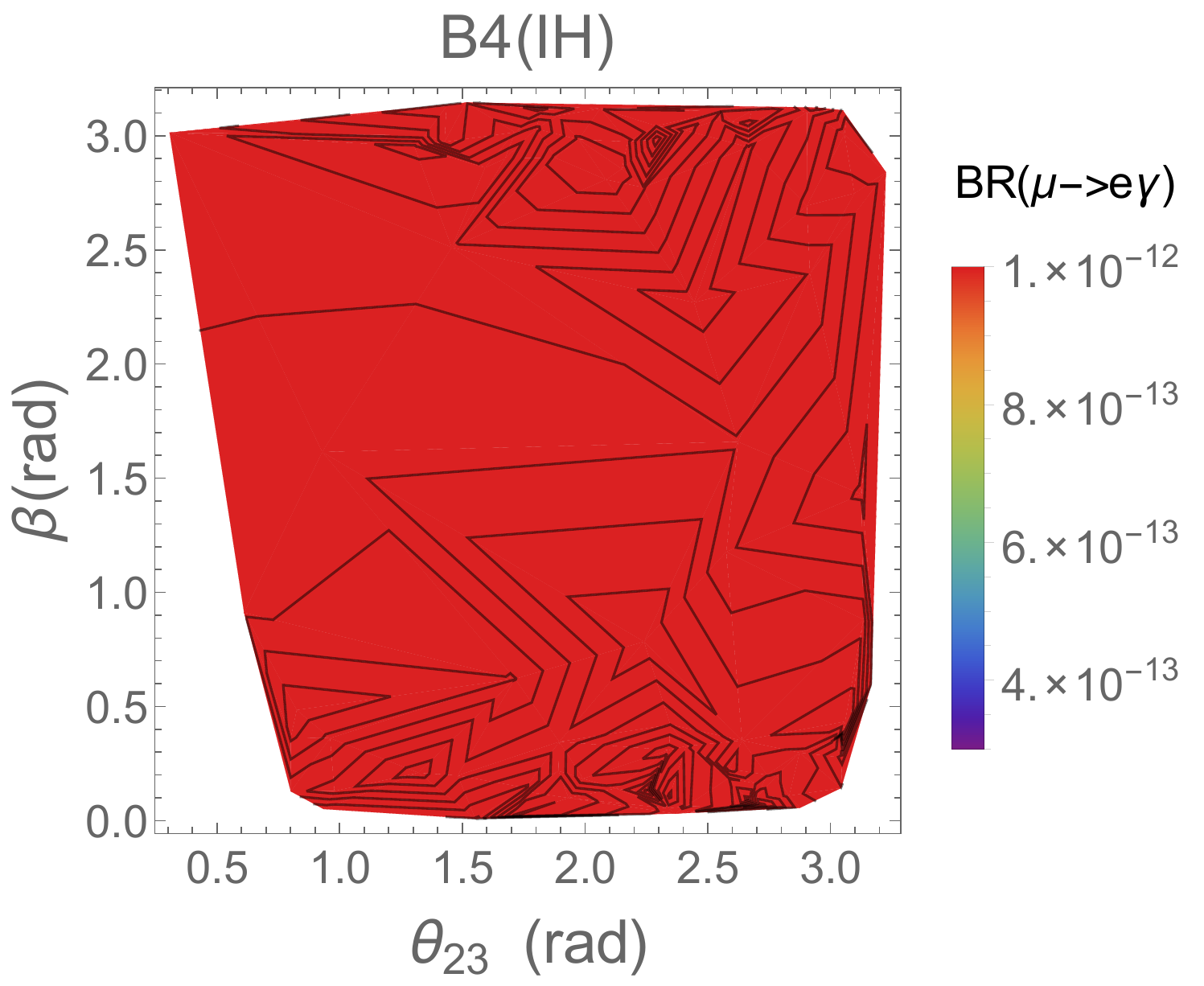}
	
	\caption{Atmospheric mixing angle, $\theta_{23}$ Vs Majorana phase $\alpha$ (for IH) with BR for $\rm \mu\rightarrow e\gamma $ as the contour where $4.2\times 10^{-13}$ is the upperlimit for BR given by MEG experiment.  } \label{fig6}
\end{figure}
	\clearpage		
	
\begin{table}[h!]
	\centering
	\begin{tabular}{| c|c| c| c| c| c|}
		\hline
		Class&W (NH/IH)(eV) & X(NH/IH)(eV) & Y(NH/IH)(eV) &Z(NH/IH)(eV)\\ \hline
		A1 &	0.01-0.05/0.03-0.1&0.04-0.1/0.03-0.06 & 0.01-0.05/0.03-0.05&0.01-0.06/0.01-0.09\\ \hline
		A2	&	0.02-0.1/0.05-0.09&0.04-0.1/0.02-0.1 & 0.01-0.07/0.02-0.1&0.04-0.09/0.04-0.09\\ \hline
		B1	&	0.01-0.07/0.03-0.05&0.02-0.05/0.02-0.04 & 0.01-0.1/0.01-0.08&0.01-0.1/0.01-0.09\\ \hline
		B2&	0.03-0.05/0.02-0.07&0.01-0.04/0.01-0.08 & 0.02-0.1/0.02-0.1&0.09-0.1/0.01-0.1\\ \hline
		B3	&	0.02-0.06/0.01-0.08&0.01-0.1/0.02-0.1 & 0.02-0.09/0.02-0.1&0.02-0.07/0.02-0.08\\ \hline
		B4	&	0.02-0.07/0.02-0.08&0.01-0.09/0.03-0.1 & 0.02-0.1/0.05-0.1&0.02-0.1/0.01-0.1\\ \hline

	\end{tabular}
	\caption{The range of model parameters that satisfies the KamLAND-Zen limit of effective Majorana neutrino mass.} \label{t5}
\end{table}
\begin{table}[h!]
	\centering
	\begin{tabular}{| c|c| c| c| }
		\hline
		Class&$\alpha (rad)$ (NH/IH)& $\beta (rad)$(NH/IH) & $m_{lightest}(eV) $(NH/IH)\\ \hline
		A1 &	1.5-3.8/1.5-3.8&1.0-3.0/1.0-3.0  & 0.01-0.1/0.01-0.1\\ \hline
		A2	&	1.2-3.7/1.7-3.2&1.2-3.0/0.5-3.0  & 0.01-0.1/0.01-0.1\\ \hline
		B1	&	1.7-3.7/1.7-3.3&1.2-3.0/1.0-3.0  & 0.01-0.1/0.01-0.1\\ \hline
		B2  &	1.5-3.2/1.5-3.7&1.2-3.0/1.2-3.0 & 0.01-0.1/0.01-0.1\\ \hline
		B3	&	1.5-4.0/1.5-3.7&1.2-3.0/1.2-3.0   & 0.01-0.1/0.01-0.1\\ \hline
		B4	&	1.5-3.5/0.8-3.2&0.7-3.0/0.7-3.0  & 0.01-0.1/0.01-0.1\\ \hline

	\end{tabular}
	\caption{The range of parameters ($\alpha, \beta $ and $m_{lightest}$) that satisfies the KamLAND-Zen limit of effective Majorana neutrino mass.} \label{t5}
\end{table}
\begin{table}[h!]
	\centering
	\begin{tabular}{| c|c| c| c|}
		\hline
		Class&$\theta (rad)$ (NH/IH)& $\phi (rad)$(NH/IH) \\ \hline
		A1 &	0.05-0.35/0.05-0.3&1.57-1.58/1.57-1.60 \\ \hline
		A2	&	0.05-0.35/0.05-0.3&1.62-1.66/1.62-1.66\\ \hline
		B1	&	0.05-0.35/0.05-0.3&1.57-1.65/1.57-1.60 \\ \hline
		B2  &	0.05-0.3/0.05-0.35&1.62-1.66/1.57-1.65 \\ \hline
		B3	&	0.05-0.35/0.05-0.3&1.62-1.66/1.62-1.66\\ \hline
		B4	&	0.35-0.55/0.45-0.55&1.57-1.60/1.63-1.66\\ \hline

	\end{tabular}
	\caption{The range of TM mixing parameters ($\theta$ and $\phi $)  that satisfies the KamLAND-Zen limit of effective Majorana neutrino mass.} \label{t5}
\end{table}

\subsection{COLLIDER SIGNATURES}

Physics at TeV scale has obtained great importance owing to the fact that it can be probed at the colliders.  Characteristic signatures of the LRSM (which is the framework of our concern)  at the hadron collider experiments like LHC emerges from the production and decay of triply and doubly charged scalars of the scalar quadruplet. In TeV scale  LRSM, the presence of RH gauge interactions as well as the mixing between the heavy and light neutrinos lead via production of the RH gauge boson, $W_R$ to significant signal strength for the $l^{\pm}l^{\pm}jj$ channel. In the colliders $W_R$ could be produced
through Drell-Yan, which decays to RH neutrino and a charged lepton. The RH neutrino (which are Majorana particles) can further decay to charged leptons/antileptons and jets. With negligible mixing between heavy and light neutrinos as well as left and right W bosons, both $W_R$ and $N_R$ couple through RH currents. Several constraints have been put forwarded on the mass of the RH gauge boson, $W_R$, the breaking scale of LRSM based on low energy processes like leptogenesis, supersymmetry, neutrinoless double beta decay etc. Most stringent experimental constraints on the masses of $W_R$, $M_N$ in MLRSM as explained in \cite{mitra2016neutrino} are provided by  $l^{\pm}l^{\pm}jj$ searches in ATLAS, dijet searches by ATLAS (CMS), neutral hadron transitions and search for NDBD. When the breaking scale of LRSM is low enough, LNV can be seen and hence the Majorana nature of the neutrino mass can be probed in the colliders and in future experiments in a wider range of parameter space. Since we are considering the low energy phenomenon like NDBD and LFV, we are considering the experimental bounds on these mass provided by the search for these phenomenon.
The NDBD experiments are mainly focused in determining the effective Majorana neutrino mass $<m_{\beta\beta}>$ which is related to the observed NDBD lifetime as,
\begin{equation}
\frac{1}{{T_{\frac{1}{2}}}^{0\nu}}=G^{0\nu}(Q,Z){\left|M^{0\nu}\right|}^2\frac{{\left|m_{\beta\beta}\right|}^2}{{m_e}^2},
\end{equation}
where the terms $G_{0\nu}$, $M_\nu$ and $m_e$ represents the phase space factor, the nuclear matrix element (NME) and the electron mass respectively. $\Gamma$
represents the decay
width for $0\nu\beta\beta$ decay process. The best lower limits on the NDBD half life has been obtained for the isotopes Ge-76, Te-130, Xe-136 in notable experiments like GERDA-II, CUORE, KamLAND-Zen respectively. The non observation of NDBD constraints the masses of  $W_R$ and $N_R$ as, $\sum_i \frac{{Y_{ei}}^2}{M_i {M_{W_R}}^4}$ 	$\leq$ (0.082-0.076) $TeV^{-5}$ using 90$\%$ CL from the limit propounded by KamLAND-Zen $ \rm T_{1/2}^{0\nu}>1.07\times 10^{26}$ which corresponds to an effective mass of $|<m_{eff}>|<(0.061-0.065)eV$ \cite{gando2016search} where the range corresponds to the uncertainities in the NMEs of the relevant process. For $M_{W_R}$ of 3 (5 TeV), the mass of the RH $\nu$ $\geq$ 150-162 GeV (19.5-21)GeV. Again, Tello et al. \cite{tello2011left} found the lower bound on mass of ${\Delta_R}^{++}$ to be

\begin{equation}
M_{\Delta_R}^{++}\geq 500 \left(\frac{3.5 TeV}{M_{W_R}}\right)^2 \times \sqrt{\frac{M_N}{3 TeV}}
\end{equation}

Considering these experimental bounds in mind, we have considered the mass of $W_R$ as 3.5 TeV in accordance with the collider probes  and the other heavy particles of the order of TeV.

\section{CONCLUSION}
The importance of texture zero neutrino mass and its phenomenological consequence has gained utmost significance in present day research. In this context two zero texture neutrino mass matrices are more relevant as they  provides the minimal free parameters for precise study. We have performed a systematic study of the Majorana neutrino mass matrix with two independent zeros. As has been pointed out in several earlier works that seven out of fifteen patterns namely (A1, A2, B1-B4, C1) can survive the current experimental data at 3$\sigma$ level.
We tried to study the constraints of the allowed patterns of texture zero neutrino  mass matrices in the framework of LRSM from low energy phenomenon like NDBD and LFV. We have shown that one can obtain the desired two zero texture mass matrices by implementing a abelian discrete symmetric group $Z8\times Z2$ in the framework of left-right symmetric model. The two zero textured neutrino mass matrix in our case is able to explain NDBD with the effective Majorana mass within the experimental limit propounded by experiment (KamLAND-Zen). However all the different allowed classes of two zero textures shows different results for different neutrino mass hierarchies. Based on our results, having done a careful comparison of the plots obtained for different classes of two zero textures, it is seen that none of the cases totally disallows NDBD as far as the KamLAND-Zen limit is concerned irrespective of the mass hierarchies. However the allowed range of the parameter space is constrained for the allowed experimental bounds of effective Majorana neutrino mass.   We have considered six different allowed  classes of two zero texture neutrino mass matrices which satisfies TM mixing in our case. Again we have done an analysis of the model parameters (W, X, Y, Z) in our case which are heavily constrained for a very limited parameter space for some classes, specifically B1 (IH), B2 (NH) for some combinations of the model parameters which has been explained in the numerical analysis. Thus we can say that the contributions from the type II SS in NDBD is relatively less for these classes. Interestingly the present results ruled out B4 class (for both NH and IH)  and A1(IH), B1(IH), B3(IH), A2(NH) classes of two texture zero neutrino mass in explaining the  experimentally allowed regions of charged lepton flavour violation whereas only the class B2 is giving  results within bounds for both the mass hierarchies. Again, the Majorana phases $\alpha$ and $\beta$ are also constrained from both NDBD and LFV point of view.  However, the sensitivity of NDBD experiments to the effective mass governing NDBD will probably reach around 0.05 eV in  future experiments which might exclude or marginally allow some of the two zero texture patterns in nearby future. Notwithstanding, an indepth study of the texture zero classes considering all the model parameters and its implications for NDBD, LFV could be done for an even more strong conclusion.\\\\

\acknowledgments
The work of MKD is supported by the Department of Science and Technology, Government of India under the project no. EMR/2017/001436. The authors would like to thank Debasish Borah, IIT Guwahati, for some useful comments  regarding the plots.
\bibliographystyle{JHEP}
\bibliography{xyz}

\end{document}